# Monolayer van der Waals kagome lattice for high-temperature flat-band physics

Hyungkun Bong[1*], Choongjae Won[2*], Seongyong Lee[2,3*], Soomin Oh[4*], Phuong X. Nguyen[5], Raghav Chaturvedi[5], Sehyun Oh[1], Jiyong Han[1], Paul Malinowski[5], Chad Mowers[5], Jimin Kim[2,6], Sae Hee Ryu[7,8], Hyunjong Lee[1], Jinhyeok Choi[1], Sangsu Yer[4], Chris Jozwiak[7], Aaron Bostwick[7], Eli Rotenberg[7], Craig M. Polley[9], Matthew D. Watson[10], Timur K. Kim[10], Heejun Yang[4], Jieun Lee[1], Kin Fai Mak[5], Jie Shan[5], Jae-Hoon Park[2,3†], Gil Young Cho[4,11†], Kyle M. Shen[5,12†] & Mingu Kang[1,2,5,12†]

[1]Department of Physics and Astronomy, Seoul National University, Seoul 08826, Republic of Korea.

[2]Max Planck POSTECH/Korea Research Initiative, Center for Complex Phase of Materials, Pohang 37673, Republic of Korea.

[3]Department of Physics, Pohang University of Science and Technology, Pohang 37673, Republic of Korea.

[4]Department of Physics, Korea Advanced Institute of Science and Technology, Daejeon 34141, Republic of Korea.

[5]Laboratory of Atomic and Solid State Physics, Cornell University, Ithaca, New York 14853, USA.

[6]Pohang Accelerator Laboratory, Pohang University of Science and Technology, Pohang 37673, Republic of Korea.

[7]Advanced Light Source, Lawrence Berkeley National Laboratory, Berkeley, California 94720, USA.

[8]Multipurpose Synchrotron Radiation Construction Project, Korea Basic Science Institute, Cheongju 28119, Republic of Korea.

[9]MAX IV Laboratory, Lund University, Lund SE-221 00, Sweden.

[10]Diamond Light Source, Harwell Science and Innovation Campus, Didcot, OX11 0DE, UK.

[11]Center for Artificial Low Dimensional Electronic Systems, Institute for Basic Science, Pohang 37673, Korea.

[12]Kavli Institute at Cornell for Nanoscale Science, Cornell University, Ithaca, New York 14853, USA.

[*]These authors contributed equality to the work.

[†]Corresponding author. Email: jhp@postech.ac.kr, gilyoungcho@kaist.ac.kr, kmshen@cornell.edu, iordia@snu.ac.kr

**Kagome lattice materials host a remarkably rich landscape of novel quantum phenomena, all arising from the intrinsic interplay of topology, correlation, and frustration[1–3]. While previous studies have been mostly limited to bulk three-dimensional (3D) kagome crystals, the realization of a true two-dimensional (2D) kagome system has remained a long-standing goal, to unlock fundamentally new physics[4–6] and engineering pathways[7,8] inaccessible in 3D systems. In this work, we establish the first ideal 2D monolayer kagome lattices by leveraging a novel and versatile van der Waals (vdW) kagome family, $A_2T_3X_4$ ($A$ = K, Rb, Cs; $T$ = Ni, Pd, Pt; $X$ = S, Se). Our nano-focused angle-resolved photoemission spectroscopy (nano-ARPES) directly characterizes the electronic structure of $A_2T_3X_4$ down to the monolayer limit, notably revealing an isolated kagome-derived flat band just below the Fermi level. Via gate-engineering of 2D $A_2T_3X_4$, the kagome flat band can be further tuned to the Fermi level, allowing direct access to the partially filled flat band physics and establishing a qualitatively new 2D flat band platform. Compared to existing Moiré-based artificial flat band systems[9], this vdW kagome-based intrinsic flat band system realizes more than 220 times higher flat band electron density, opening an entirely new density and energy regime for exploring 2D flat band physics[10]. Harnessing this high-electron-density flat band in 2D vdW kagome system provides a clear pathway to scale emergent flat band phenomena – such as fractional quantum anomalous Hall states – to far higher temperatures than currently achievable, as long-envisioned in theory[4] and further supported by our many-body calculations.**

Kagome lattice systems have rapidly emerged as one of the most fertile frontiers in quantum materials research[1–3]. This seemingly simple model lattice geometry dictates a remarkably rich landscape of quantum behavior, including Dirac fermions, van Hove singularities, and flat bands in momentum-space, as well as frustrated magnetism, Kane-Mele spin-orbit coupling, and long-range entanglement in real-space. This synergistic convergence of singular electronic structures and quantum interactions in kagome lattice has proven to yield an exceptional diversity of emergent phenomena, most notably, time-reversal symmetry breaking topological orders in Mn-, Fe-, and Co-based kagome systems[11–13]; coexistence of charge order, nematicity, and superconductivity in V-based kagome systems[14,15]; and correlation-induced non-Fermi liquid behavior and pseudogap in Ni- and Cr-based kagome systems[16,17]. These emergent phenomena in kagome materials are also deeply connected to other broader themes in condensed matter physics, such as high-temperature superconductivity and fractionalized excitations, thereby advancing our fundamental understanding of quantum states of matter and further guiding their on-demand design.

The next key challenge for this emerging research field is the realization of a genuine 2D kagome lattice, with studies so far remaining mostly limited to bulk 3D kagome systems[11–17]. While 3D kagome systems have suffered from significant interlayer hybridization and complex multiorbital band structures (Fig. 1a,b), a true 2D kagome lattice would isolate the canonical single-orbital kagome bands, free from these complications (Fig. 1c). Such an isolated kagome dispersion is critical to realize the long-envisioned physics of an ideal kagome system, including quantized topological transport[18], maximal anomalous Hall angles[19], and Laughlin-type fractional ground states arising from an intrinsic topological flat band[4]. Moreover, a 2D kagome lattice would unlock fundamentally new engineering opportunities – electrostatic gating, heterostructure design, interfacial proximity coupling, and strain- or twist-angle control – that would greatly expand the scope of kagome lattice physics[7,8]. For instance, twisted-angle-engineered kagome bilayers have been theoretically proposed to realize higher-order topology and quasicrystalline electronic structures[5,6], but such architectures are accessible only through a true 2D kagome platform. Despite this promise and long-standing efforts, however, the realization of an ideal 2D kagome lattice still remains an outstanding challenge. The main obstacle here arises from the fact that nearly all kagome materials discovered to date exhibit either metallic or ionic bonding between kagome layers (Fig. 1a,b), making mechanical exfoliation or layer-by-layer growth of their 2D form fundamentally infeasible.

In this work, we overcome these limitations and establish the first ideal 2D monolayer kagome system with a model kagome band manifold near the Fermi level (Fig. 1c). Our work is enabled by a newly identified van der Waals (vdW) kagome family, $A_2T_3X_4$ ($A$ = K, Rb, Cs; $T$ = Ni, Pd, Pt; $X$ = S, Se), whose unique vdW nature allows the long-sought realization of an ideal 2D kagome lattice via straightforward mechanical exfoliation. Using nano-focused angle-resolved photoemission spectroscopy (nano-ARPES), we directly map the electronic dispersion of $A_2T_3X_4$ down to the monolayer limit, notably revealing an isolated kagome-derived flat band just below the Fermi level. This kagome flat band in 2D $A_2T_3X_4$ can be further gate-engineered to the Fermi level and driven into the partially filled regime, providing an unprecedented platform for exploring long-envisioned lattice-driven flat band phenomena, including flat band magnetism[20], Wigner crystallization[21], correlated topological orders[22], and fractional quantum anomalous Hall states[4]. We emphasize that this *intrinsic* flat band realized in 2D vdW kagome systems hosts more than two orders of magnitude higher carrier density ($n \approx 1.0$ nm$^{-2}$) compared to *artificial* flat bands in Moiré superlattices ($n \approx 0.01$ nm$^{-2}$)[9], suggesting that such flat band phenomena might be accessible at substantially enhanced energy and temperature regimes[10]. Further supporting this important prospect, our many-body calculations uncover correlated topological phases in monolayer $A_2T_3X_4$ at temperatures an order of magnitude higher than currently achievable in Moiré systems[23–25], collectively underscoring the implication of the 2D vdW kagome system for high-temperature flat band physics.

## Bulk vdW kagome crystals and exfoliated 2D kagome flakes

We identified the vdW kagome system $A_2T_3X_4$ through topological quantum chemistry and high-throughput materials screening[26,27]. $A_2T_3X_4$ constitutes a broad materials family, in which one can independently tune the alkali metal site $A$ (K, Rb, Cs), transition metal site $T$ (3*d*-Ni, 4*d*-Pd, 5*d*-Pt), and chalcogen site $X$ (S, Se)[28–34]. Among the six possible combinations of transition metal $T$ and chalcogen $X$, we successfully synthesized five compounds – $Cs_2Ni_3S_4$, $Cs_2Ni_3Se_4$, $(Rb,Cs)_2Pd_3S_4$, $(Rb,Cs)_2Pd_3Se_4$, and $Cs_2Pt_3S_4$ – using a flux growth method (Fig. 1d and Methods). A key structural feature of $A_2T_3X_4$ is the kagome lattice formed by the $T$ atoms, with each $T$ site coordinated by a low-symmetry planar arrangement of $X$ atoms (Fig. 1c, see also Extended Data

Fig. 1). This coordination imposes a strong crystal field splitting at the kagome $T$ site, effectively isolating a single $d_{xy}$ orbital degrees of freedom near the Fermi level and ultimately realizing a model single-orbital kagome dispersion, as highlighted in Fig. 1c. The Fermi level lies between the fully filled $d_{xy}$ orbital and the empty $d_{zx}$ orbital, rendering the system a semiconductor with a bandgap in the visible light range. The color of the crystals varies from gold ($Cs_2Ni_3Se_4$), olive gold ($Cs_2Ni_3S_4$), teal ($(Rb,Cs)_2Pd_3Se_4$), dark magenta ($(Rb,Cs)_2Pd_3S_4$) to silver ($Cs_2Pt_3S_4$), indicating that the electronic structure and optical band gap are widely tunable by varying the transition metal or chalcogen site, while the alkali metal site plays a marginal role[35].

We now highlight the realization of a 2D kagome system by mechanical exfoliation of bulk $A_2T_3X_4$. Representative optical images of exfoliated mono- to penta-layer $Cs_2Pd_3S_4$ on $SiO_2$/Si substrates are shown in Fig. 1e. The exfoliated flakes notably exhibit large lateral dimensions of 10–30 μm, comparable to other well-established vdW systems[7]. The thickness of 1–5 layer flakes can be reliably determined by quantized optical contrast (Fig. 1f), and atomic force microscopy measurements further confirm the single-kagome-layer thickness of ≈ 0.7 nm (Fig. 1g), consistent with the interlayer spacing in $Cs_2Pd_3S_4$. As demonstrated in Extended Data Fig. 2, the exfoliated $Cs_2Pd_3S_4$ flakes remain stable under a standard Ar glovebox atmosphere for at least 60 days and withstand temperatures up to 300 ºC. The ease of exfoliation and high environmental stability makes the $A_2T_3X_4$ series an ideal vdW kagome platform, enabling the exploration of kagome lattice physics in the genuine 2D limit for the first time. These properties sharply distinguish $A_2T_3X_4$ from previously considered vdW kagome candidates, which either exhibit significant lattice distortion in the kagome plane, resulting in low-energy physics dominated by localized molecular-orbital states (e.g., $Nb_3(Cl,Br,I)_8$)[36,37], or fail to be reliably exfoliated down to the 2D limit due to inherent structural interlocking between adjacent vdW layers (e.g., $Pd_3P_2S_8$)[38] (see Supplementary Section 1,2 for detailed discussion).

**Electronic structure of bulk vdW kagome series**

Having established $A_2T_3X_4$ as an ideal 2D kagome system, we now probe its intrinsic electronic structure using ARPES. We first discuss the electronic structure of bulk $A_2T_3X_4$ in Fig. 2, followed by the nano-ARPES characterization of 2D $A_2T_3X_4$ devices in Fig. 3. Fig. 2c,d display the bulk ARPES spectra of the representative $Rb_2Pd_3S_4$ along the $\bar{\Gamma}$–$\bar{M}$ and $\bar{\Gamma}$–$\bar{K}$ high-symmetry surface Brillouin zone directions, respectively. Corresponding DFT band structures, calculated using the hybrid functional, are also displayed in Fig. 2f,g. The most important feature of the $A_2T_3X_4$ electronic structure is a set of three bands located just below the Fermi level, highlighted by red and black solid lines in Fig. 2f,g. The topmost band (red) realizes a quasi-flat band, while the two lower bands (black) form a Dirac point. These are the band manifold from $T$-$d_{xy}$ orbitals, which together provides a model realization of the canonical kagome electronic structure, as discussed in Fig. 1c (see Supplementary Section 3 for details of orbital character assignment). We emphasize that the quasi-flat band in $Rb_2Pd_3S_4$ remains almost completely dispersionless along $\bar{\Gamma}$–$\bar{M}$ ($W_{flat}^{\Gamma-M}$ < 70 meV, Fig. 2c) and acquires only a small dispersion near $\bar{K}$ ($W_{flat}^{\Gamma-K}$ < 180 meV, Fig. 2d), realizing one of the flattest dispersions ever observed in kagome systems (Extended Data Fig. 3). Notably, due to their in-plane orbital characters, this $d_{xy}$ flat band also remains dispersionless along the out-of-plane momentum direction, as demonstrated by our photon-energy-dependent ARPES measurements and DFT calculations (Fig. 2e,h). As discussed above, these $d_{xy}$ kagome bands are well-separated from other $d$-orbital states due to the strong crystal field in $A_2T_3X_4$ (Fig. 1c), offering an ideal platform to investigate clean single-orbital kagome lattice physics. Such an isolated model kagome dispersion – in particular, the isolated kagome flat band – has not been achieved in previously studied metallic or ionic kagome systems (Fig. 1a,b)[11–17], and represents a unique material-level advancement in this vdW kagome series (Fig. 1c).

Notably, our comprehensive ARPES measurements across the $A_2T_3X_4$ series, shown in Extended Data Fig. 4, further reveal that these $d_{xy}$ kagome dispersions are highly tunable by varying $T$ (Ni, Pd) and $X$ (S, Se) compositions. To illustrate this, we summarized the evolution of key band parameters as a function of chemical composition in Fig. 2i–l. First, the *overall* $d_{xy}$ *bandwidth* is larger for $T$ = 4$d$-Pd than $T$ = 3$d$-Ni compounds, reflecting stronger electronic hopping in 4$d$ orbitals (Fig. 2i). In contrast, when focusing specifically on the flat band, the *flatness of the quasi-flat band* is more strongly affected by the chalcogen $X$ atoms, which set the ratio between nearest-neighbor and longer-range hoppings (Fig. 2j). Specifically, the quasi-flat band is significantly flatter for $X$ = S compounds compared to $X$ = Se counterparts, and the flattest bandwidth of $W_{flat} \approx$ 180 meV is achieved in $A_2Pd_3S_4$. Moreover, the *spin-orbit coupling gap* at the band degeneracy points is larger in 4$d$-Pd compounds (Fig. 2k), whereas the *degree of correlation* is stronger for 3$d$-Ni compounds, as estimated from the renormalization factor between experimental and DFT band structures (Fig. 2l). Altogether, these surveys demonstrate that not only the kagome dispersion but also the strengths of spin-orbit coupling and electronic correlations can be systematically engineered within this versatile $A_2T_3X_4$ family. Below, we focus on the case of $A_2Pd_3S_4$, featuring the most quenched flat band state among the series.

**Electronic structure of 2D vdW kagome devices**

With the electronic structure of bulk $A_2T_3X_4$ fully characterized, we next examine how the electronic structure evolves in the ultrathin, monolayer limit of $A_2T_3X_4$ using nano-ARPES. For the nano-ARPES measurements, we first fabricate a vdW heterostructure device [graphene / $A_2T_3X_4$ / graphite], in which a 2D $A_2T_3X_4$ flake was encapsulated between top graphene and bottom graphite layer (see Fig. 3a for schematic illustration). Here, the bottom graphite provides an atomically flat support that enhances the crystallinity of the ultrathin kagome flake placed above, while the top graphene encapsulates the kagome flake, protecting it against environmental degradation. We refer to Extended Data Fig. 5,6 and Supplementary Section 3–5 for further details of the nano-ARPES measurements.

Figure 3b–e display the optical image of our graphene/$A_2Pd_3S_4$/graphite heterostructure (Fig. 3b), along with nano-ARPES spatial maps highlighting each layer *in situ* (Fig. 3c–e). Specifically, angle-integrated valence band photoemission intensity (Fig. 3c), the S-2*p* core level (Fig. 3d), and graphene Dirac band intensity (Fig. 3e) were selected to visualize the bottom graphite (white solid lines), $A_2Pd_3S_4$ flake (magenta solid lines), and top graphene (yellow solid lines), respectively. These *in situ* maps reveal two $A_2Pd_3S_4$ flakes on our heterostructure – one covered by graphene and the other exposed – as marked by the magenta and green outlines in Fig. 3b–e. By comparing the S-2$p_{3/2,1/2}$ core level spectra of the two flakes (inset in Fig. 3d), we find that only the encapsulated flake exhibits a sharp doublet spectrum, while the exposed flake displays much broader spectrum with binding energy shift, indicating clear signs of degradation. This spectro-microscopic information thus directly demonstrates the high stability of 2D $A_2T_3X_4$ once encapsulated in a vdW heterostructure, which is critical for facilitating future investigations of this vdW kagome family.

By focusing the nano-ARPES beamspot onto the encapsulated $A_2Pd_3S_4$ flake, we mapped the Fermi surface and constant energy contours of the graphene/$A_2Pd_3S_4$/graphite heterotrilayer, as shown in Fig. 3f,g. At the Fermi level, two sets of Dirac points forming two large hexagonal Brillouin zones can be clearly identified (see yellow and gray hexagons in Fig. 3f), which correspond to Dirac electrons from the top graphene and bottom graphite layers, respectively.

Upon lowering the energy, the Dirac points evolve into trigonally wrapped hole pockets, and at the same time, new photoelectron intensity with a much smaller Brillouin zone periodicity emerges (magenta hexagons in Fig. 3g), highlighting electronic states from $A_2Pd_3S_4$. Owing to the large mismatch in the Brillouin zone sizes, measurements of the $A_2Pd_3S_4$ bands were not affected by the presence of graphene and graphite bands (see Extended Data Fig. 6). From the relative orientation of the Brillouin zones, we determined the twist angle between graphene and $A_2Pd_3S_4$ to be 28.9º, and that between $A_2Pd_3S_4$ and graphite to be 20.2º for this specific heterotrilayer.

By fabricating multiple such heterostructures with varying $A_2Pd_3S_4$ thickness (Extended Data Fig. 7), we systematically tracked the evolution of the electronic structure of $A_2Pd_3S_4$ from 3D bulk to the 2D monolayer limit, as shown in Fig. 3h,i. Given the minimal variation of the electronic structure across different *A*-site elements[35], we compile data from both $Rb_2Pd_3S_4$ (bulk and 8 layer) and $Cs_2Pd_3S_4$ (4, 2, 1 layers). The spectrum quality varies across devices; nevertheless, sharp non-dispersive flat bands (Fig. 3h) as well as linearly dispersing Dirac bands (Fig. 3i) are clearly resolved down to the true 2D, monolayer kagome lattice limit. The overall $d_{xy}$ dispersions exhibit minimal dependence on layer thickness, consistent with their in-plane orbital character and weak interlayer coupling in vdW $A_2T_3X_4$. However, the electronic structure in the monolayer limit displays several distinctive features: the $d_{xy}$ kagome bands become not only dramatically sharper, potentially reflecting the quenching of interlayer effects, but also emerge much closer to the Fermi level, lying within only 420 meV. This shift toward Fermi level signifies strong charge transfer from the monolayer $A_2T_3X_4$ to the adjacent graphene/graphite layers, suggesting the sensitivity of the Fermi level in 2D $A_2T_3X_4$ to local environments, typical of 2D vdW materials[7,8]. Altogether, our nano-ARPES experiments provide the first direct electronic structure characterization of a 2D vdW kagome system, revealing an ideal kagome dispersion, in particular the kagome-derived flat band, robust down to the monolayer limit. Our investigations therefore establish 2D $A_2T_3X_4$ not only as a long-sought 2D kagome lattice, but also as a fundamentally new 2D flat band platform distinct from existing Moiré-based systems, as discussed further below.

**Gate-tunable flat band states and potential for high-temperature flat band phenomena**

The isolated flat band realized in monolayer $A_2Pd_3S_4$ highlights the important potential of vdW kagome systems as a qualitatively new 2D flat band platform. Until now, the investigation of 2D flat band physics has mostly relied on Moiré superlattice systems, in which the flat band in momentum-space is artificially generated by large Moiré supercells in real-space (Fig. 5c)[9]. This approach, while efficiently converting non-flat band systems into flat-band systems, however provides only one flat band electron per large Moiré supercell ($a \approx 10$ nm), resulting in an extremely low carrier density ($n \approx 0.01$ nm$^{-2}$). This dilute flat band electron density is a fundamental limitation of the Moiré systems, that inevitably constrains the energy scale of flat band phenomena to very low temperatures (Fig. 5b)[10]. In a heuristic picture, Moiré superlattices ($a \approx 10$ nm) simulate physics between true crystalline solids ($a \approx 1$ nm) and ultracold optical lattices ($a \approx 1000$ nm), possessing correspondingly reduced energy and temperature scales (see also Supplementary Section 6). In contrast, the flat band in 2D vdW kagome system arises from a fundamentally different mechanism – destructive interference of electron hopping in a frustrated geometry – without need for such superlattice potential (Fig. 5a)[3]. This intrinsic mechanism provides one flat band electron per unit cell ($a \approx 1$ nm), giving rise to more than two orders of magnitude higher flat band electron density compared to the Moiré approaches ($n \approx 1$ nm$^{-2}$). This dramatically enhanced flat band electron density in 2D kagome system opens a fundamentally new physical regime for exploring 2D flat band phenomena, with potential to elevate flat band physics to far higher energy and temperature ranges[10].

To fully establish 2D vdW kagome lattice as a realistic new flat band platform, it is essential to engineer flat band in $A_2T_3X_4$ to the Fermi level and access the partially filled regime of the kagome flat band. While such control has remained fundamentally challenging in previously studied 3D kagome systems, the 2D nature of $A_2T_3X_4$ offers multiple direct pathways to tune the flat band, for example, via heterostructure and electrostatic gate engineering. To experimentally demonstrate the tunability of the flat band in $A_2T_3X_4$, we fabricate 2D kagome field effect devices and perform systematic gate-dependent spectroscopic and transport measurements, as shown in Fig. 4 (see also Supplementary Section 7 for details). First, in a single-gated device configuration with an exposed top surface, our *in operando* nano-ARPES experiments directly visualize the tunability of the flat band over a 390 meV range upon applying the gate voltage $V_G$ from −7.5 V to 11 V (Fig. 4a,b). Although the applicable gate range is practically constrained in nano-ARPES settings (see Methods), we observe that the flat band energy noticeably saturates beyond $V_G < -2.5$ V, suggesting that the flat band top has already reached the Fermi level and has started to be carrier-doped[39,40]. To complement this spectroscopic observation, we further fabricate dual-gated two-terminal devices to directly characterize the transport response of the 2D vdW kagome lattice upon gating (Fig. 4c). At neutral and positive gate voltages (0 V < $V_{TG}$ < 5 V), the device remains highly insulating (Fig. 4d) and the temperature-dependence also displays insulating response (Fig. 4e,h), consistent with a fully filled flat band state and a large band gap in pristine $A_2T_3X_4$. By contrast, application of negative gate voltage (−2.5 V < $V_{TG}$ < 0 V) drives a clear insulator-to-metal transition across the band edge, characterized by more than three orders of magnitude reduction in channel resistance (Fig. 4d) and the development of metallic behavior in the temperature-dependence (Fig. 4e,h). This emergence of robust metallic transport in originally insulating $A_2T_3X_4$ unambiguously demonstrates successful tuning of the kagome flat band to the Fermi level via gate-induced carrier doping, and direct probes the response from partially filled kagome flat band electrons. Interestingly, upon applying further negative gate (−5.5 V < $V_{TG}$ < −2.5 V), the high-temperature resistance continues to decrease as expected from increasing hole carriers (Fig. 4f), but the overall temperature-dependence develops insulating character (Fig. 4g,i), realizing a new insulator state within the partially filled flat band state. As trivial scenarios such as disorder-dominated insulating state are unlikely to occur beyond the band-metal regime, this insulating state may potentially indicate the onset of many-body effects in the 2D kagome flat band system, motivating future optical and tunneling spectroscopy studies to determine its precise nature. Altogether, our gate-dependent measurements firmly establish gate-tuned 2D vdW kagome lattices as a new class of partially-filled flat band system, providing fundamentally new opportunities to study 2D flat band physics beyond existing Moiré-based approaches. This invites for future experimental efforts – including systematic surveys of contact metals to optimize contact resistance; fabricating more sophisticated Hall-bar geometries; access to higher carrier densities through ionic liquid gating or chemical doping; and complementary optical and tunneling spectroscopy investigations for identifying the microscopic nature of correlated phases, etc – to ultimately map and understand the full landscape of flat band phenomena in the 2D vdW kagome system.

While establishing the full landscape of flat band phenomena in the 2D vdW kagome system will require further experimental investigations, the demonstrated access to partially filled flat band states in $A_2T_3X_4$ motivates us to explore, from a theoretical perspective, the potential landscape of flat band phenomena realizable in the 2D vdW kagome platform. In particular, to explicitly compare the energy scales of flat band phenomena in kagome- and Moiré-based flat band systems, we investigate possible fractional quantum anomalous Hall ground states in $A_2Pd_3S_4$, which is recently observed in Moiré flat bands (twisted-bilayer $MoTe_2$) below 3 K[23–25]. First, we emphasize that the flat band in 2D $A_2Pd_3S_4$ is sufficiently flat ($W_{flat}{}^{2D} \approx 0.13$ eV) compared to Hubbard interaction energy scale of transition metal *d*-orbitals ($U \approx$ 1–6 eV), giving rise to a cascade of correlation-driven electronic orders in the mean-field phase diagram of 2D $A_2Pd_3S_4$, including triangular antiferromagnetism, spin-charge-stripe phases, topological noncoplanar spin states, and flat-band ferromagnetism (Fig. 5f). The largest portion of the phase diagram is populated by the ferromagnetic phase, a well-known exact ground state of the kagome Hubbard model[20]. In this time-reversal symmetry broken phase, the flat band in $A_2Pd_3S_4$ possesses a nonzero Chern number ($C$ = 1) and nontrivial quantum geometry (Fig. 5g,h), which is another defining characteristic of kagome flat bands. Notably, the quantum geometry metric $\frac{A_{BZ}}{2\pi}\left(\mathrm{Tr}\, g_{\mathrm{k}} - \Omega_{\mathrm{k}}\right)$ in the $A_2Pd_3S_4$ flat band remains fairly uniform across the Brillouin zone, mimicking the ideal lowest Landau level condition ($A_{BZ}$ the Brillouin zone area, $g_{\mathrm{k}}$ quantum metric, and $\Omega_{\mathrm{k}}$ Berry curvature)[41]. Accordingly, our exact diagonalization uncovers that the many-body spectra of $A_2Pd_3S_4$ also mirror those of the lowest Landau level, and notably stabilize robust 1/3 fractional quantum anomalous Hall states at fractional fillings, as comprehensively demonstrated by a threefold quasi-degenerate ground state in Fig. 5i, spectral flow consistent with the Laughlin-type fractional state in Fig. 5j, and a calculated many-body Chern number of 1/3. We also find that the fractional ground state in $A_2Pd_3S_4$ remains robust against band-mixing effects and under different model parametrizations, as further discussed in Supplementary Section 8. Most importantly, these fractional ground states in $A_2Pd_3S_4$ are characterized by large energy gaps, 14–20 meV for $Rb_2Pd_3S_4$ and 29 meV for $Cs_2Pd_3S_4$ (Extended Data Fig. 9). These energy gaps in 14–29 meV range far exceed 1–5 meV gaps in existing Moiré-based flat band systems (Fig. 5b)[42]. This explicit energy scale comparison thus supports that fractional quantum anomalous Hall states, as well as other emergent flat band phenomena, might be realized at much higher energy and temperature scales by harnessing the high-electron-density flat band realized in 2D vdW kagome system.

Altogether, by leveraging the highly versatile vdW kagome series $A_2T_3X_4$, we established the first genuine 2D kagome system hosting an ideal kagome flat band. This achievement removes the long-standing materials bottleneck that has limited kagome physics to 3D systems, and unlocks broad new opportunities to expand kagome lattice physics via electrostatic gating, heterostructure engineering, proximity effect, and strain and twist-angle engineering. Most importantly, the high-electron-density flat band realized in 2D vdW kagome systems charts a realistic pathway to scale emergent flat band phenomena to far higher energy and temperature regimes than so far achievable in current artificial flat band platforms. The prospect to realize exotic flat band-based quantum matter phases at accessible temperatures harbors profound implication for future quantum-device and quantum-information technologies.

## Methods

**Bulk crystal growth.** Single-crystals of $A_2T_3X_4$ ($A$ = K, Rb, Cs; $T$ = Ni, Pd, Pt; $X$ = S, Se) were grown using typical self-flux methods[28]. The mixture of alkali carbonates, transition metal, and chalcogenides are placed in alumina crucibles with additional alkali carbonates and chalcogenides. The crucibles were covered by lids, then heated to 850–900 °C in Ar gas condition. Then, they are slowly cooled to room-temperature. After growth, they were quickly moved to Ar-filled glove box and mechanically extracted from flux. The single-crystalline phase was confirmed by home-lab X-ray diffraction measurements.

**Exfoliation of 2D kagome lattice.** 2D $A_2T_3X_4$ flakes were obtained by mechanically exfoliating bulk $A_2T_3X_4$ onto $SiO_2$/Si substrate (275 nm oxide thickness) using adhesive tape. Flake thicknesses were determined by quantized optical contrast under an optical microscope and further confirmed by atomic force microscopy (Fig. 1f,g). The environmental stability of 2D $A_2T_3X_4$ was examined by optical microscopy and Raman spectroscopy, as detailed in Extended Data Fig. 2. Raman measurements were performed at room-temperature using a home-built setup combining a $\lambda$ = 514 nm laser (Cobolt Fandango), a Princeton Instruments TriVista spectrometer, and a liquid-nitrogen-cooled CCD (PyLoN eXcellon). We confirm that 2D $A_2T_3X_4$ remains stable under a standard Ar atmosphere, where $O_2$ and $H_2O$ partial pressures are maintained below 1 ppm. Moreover, as discussed in Fig. 3, encapsulated $A_2T_3X_4$ within vdW heterostructures remains robust under air and solvent exposure. Such high stability of 2D $A_2T_3X_4$ is a key factor enabling detailed spectroscopic and transport characterization of this 2D kagome family.

**vdW heterostructure fabrications.** Monolayer graphene, ~10 nm graphite, ~20 nm hexagonal boron nitrides (hBN), and 1–8 layers $A_2T_3X_4$ were used to assemble vdW heterostructures for nano-ARPES and transport measurements. We used all-dry-transfer method for fabrication of the heterostructure: A dome-shaped PC/PPC/PDMS stamp (PC: polycarbonate, PPC: polypropylene carbonate, PDMS: polydimethylsiloxane) was used to sequentially pick-up vdW layers for each device. During transfer, the substrate temperature was maintained at 70–85 °C while lowering the stamp and 85–100 °C while raising it. To enhance the pick-up yield of $A_2T_3X_4$ flakes, $SiO_2$/Si surface was rendered hydrophobic by substituting native silanol groups with a hydrophobic molecular layer, which effectively reduces adhesion between $A_2T_3X_4$ and $SiO_2$. After completing the stack, the assembled heterostructure was released onto a Pt/Ti electrode pre-patterned $SiO_2$/Si substrate at 200 °C. All exfoliation and stacking steps were performed in an Ar-filled glove box to minimize degradation of $A_2T_3X_4$. Once encapsulation was complete, the device was taken out of the glove box, and the stamp polymer was dissolved in chloroform for 30 min.

**Bulk electronic structure measurements.** The electronic structure of bulk $A_2T_3X_4$ was measured at two different ARPES endstations: the nanoARPES endstation of Beamline 7.0.2 (MAESTRO) at the Advanced Light Source and the A-branch endstation of the BLOCH beamline at MAX-IV. All measurements were performed at liquid helium temperature (< 20 K), under ultrahigh vacuum (< $5\times10^{-11}$ torr), and using hemispherical electron analysers (Scienta Omicron) equipped with photoelectron deflectors. The energy and momentum resolutions were better than 40 meV and 0.01 $Å^{-1}$, respectively. For each $A_2T_3X_4$ ($T$ = Ni, Pd; $X$ = S, Se), the photon energy was scanned from 60 eV to 200 eV to identify the out-of-plane Brillouin zone periodicity and high-symmetry points along $k_z$. For $A_2Pd_3S_4$, a photon energy of 98 eV was identified to lie close to $k_z \approx 0$. To optimize the photoelectron intensity of the $d_{xy}$ kagome bands of interest, we also explored different photon polarizations, experimental geometries, and Brillouin zones. The optimal photoemission matrix element was obtained under linear horizontal photon polarization (LH) and in the 2nd Brillouin zone (see also Supplementary Section 3,5 for details). Given these experimental considerations, all $A_2Pd_3S_4$ ARPES data in the Main text were obtained using LH-polarized 98 eV photons in the 2nd Brillouin zones. The corresponding ARPES data for $Rb_2Pd_3Se_4$, $Cs_2Ni_3S_4$, and $Cs_2Ni_3Se_4$ in Extended Data Fig. 4 were acquired with 110 eV, 124 eV, and 138 eV photons, respectively. It is important to mention that our initial ARPES measurements on *in situ* cleaved bulk $A_2T_3X_4$ suffered from severe charging effects, which are common in photoemission experiments on semiconducting samples. To mitigate charging, we first exfoliated $A_2T_3X_4$ onto conducting substrates (0.5 wt% Nb-$SrTiO_3$ (Shinkosha)) in a glove box environment at MAESTRO or in the loadlock chamber at BLOCH. The substrates were subsequently transferred into the ultrahigh vacuum chamber and mildly annealed at 250 °C to remove surface adsorbates. Appropriate $A_2T_3X_4$ flakes were then identified *in situ* by performing spatially-resolved ARPES measurements. Micro-focused beamspots at the two endstations – 1×1 $\mu m^2$ at MAESTRO and 5×10 $\mu m^2$ at BLOCH – were essential for this purpose. Specifically, we selected the flakes that were thin enough (< 100 nm) to suppress charging, yet thick enough (> 30 nm) to faithfully represent bulk electronic structure.

After ARPES measurements, photoelectron spatial maps were compared with optical microscope images taken *ex situ*, as shown in Fig. 2a,b. We note that similar experimental scheme was employed in previous electronic structure studies of insulating vdW crystals, such as hexagonal boron nitride, and vdW magnets, including $NiPS_3$, $MnPS_3$, and CrSBr[43–46]. We also note that this approach is effective only for flakes with sufficient thickness; measurements of few-layer flake suffer from the roughness of underlying substrates, which results in poor crystallinity of thin $A_2T_3X_4$ flakes. This highlights the necessity of the vdW heterotrilayer architecture (graphene/$A_2T_3X_4$/graphite) for measuring the electronic structure of 2D limit $A_2T_3X_4$, as discussed below.

**Device electronic structure measurements.** The nano-ARPES measurements were conducted at two different endstations: the $A_2Pd_3S_4$ devices presented in the manuscript were measured at the nanoARPES endstation of Beamline 7.0.2 (MAESTRO) at the Advanced Light Source, while the corresponding measurements on $A_2Pd_3Se_4$ devices (not shown) were measured at the nanoARPES endstation of I05 beamline at the Diamond Light Source. Capillary mirror optics were employed at both beamlines to microfocus the synchrotron radiation, achieving $1\times1$ μm$^2$ and $4\times4$ μm$^2$ spatial resolution, respectively. All measurements, except for the *in operando* measurements in Fig. 4a,b, were performed at liquid helium temperature (< 20 K), under ultrahigh vacuum (< $8\times10^{-11}$ torr), and using hemispherical electron analysers (Scienta Omicron) equipped with photoelectron deflectors. For the nano-ARPES measurements, 2D $A_2T_3X_4$ flake was encapsulated between top graphene and bottom graphite layers, forming graphene/$A_2T_3X_4$/graphite heterotrilayer structure. Such a configuration has been established to produce high quality nano-ARPES spectra of 2D materials[47–49], where the atomically flat graphite support enhances the crystallinity of the 2D flake placed above, and the graphene capping layer protects the flake from environmental degradation. Optical images of typical nano-ARPES devices are shown in Fig. 3b as well as in Extended Data Fig. 5,7. For the gate-dependent nano-ARPES experiments shown in Fig. 4a,b, a 15 nm hBN layer was inserted between $A_2T_3X_4$ and the bottom graphite, forming a graphene/$A_2T_3X_4$/hBN/graphite heterotetralayer structure (see schematics in Fig. 4a). The graphite and hBN serve as the bottom gate and gate dielectric, respectively, while carrier doping is mediated through the top graphene layer. The gate-dependent measurements were performed at room-temperature to ensure sufficient lateral conductance through the top graphene. One technical limitation of *in operando* nano-ARPES measurements is that the hBN layer develops finite conductivity upon illumination of synchrotron light, significantly reducing the gating efficiency. Together with the absence of a top gate in the device configuration, this fundamentally limits the gate-tunability of the nano-ARPES device to ~30 % compared to that achieved in standard transport device. Despite this limitation, clear gate dependent evolution and saturation of the flat band position beyond $V_G < -2.5$ V were observed, demonstrating successful tuning of the kagome flat band toward the Fermi level. All nano-ARPES samples were prepared at home institutions and transferred through air to the beamlines. Before nano-ARPES measurements, the samples were mildly annealed at 250 °C for ~6 hours to remove surface adsorbates.

**Device transport measurements.** Dual-gated two-terminal transport devices were fabricated to characterize the gate-tunability of 2D $A_2T_3X_4$ via electrical transport. The device consists of graphite/hBN top-gate; graphite electrodes; trilayer $Rb_2Pd_3Se_4$; $SiO_2$/Si bottom gate, as schematically illustrated in Fig. 4c. In 2D transport measurements, a key challenge emerges from engineering right contact electrodes that minimize the Schottky barrier. While Pt and Au electrodes result in excessively large contact resistance on $A_2T_3X_4$, graphite contacts provide moderate contact resistance, enabling initial transport characterizations of $A_2T_3X_4$. In our device configuration, the global $SiO_2$/Si bottom gate simultaneously tunes both the contact barrier and the channel carrier density, while the top gate primarily modulates the channel region. For the transport measurements in Fig. 4e–i, we fixed the bottom gate at a large negative voltage ($V_{BG} = -90$ V) to achieve good electrical contact, and tuned the top gate to control the carrier density within the channel, enabling measurements of the intrinsic gate-dependent transport response of 2D $A_2T_3X_4$. The carrier density $n$ in the channel was estimated using a simple parallel-plate capacitor model, i.e. $n = (C_{BG}V_{BG} + C_{TG}V_{TG})/e$, where $C_{BG}$ and $C_{TG}$ are the geometric capacitances per unit area for the back-gate and top-gate, respectively. These capacitances are calculated as $C_{BG}=\varepsilon_0\varepsilon_{SiO_2}/d_{SiO_2}$ and $C_{TG}=\varepsilon_0\varepsilon_{hBN}/d_{hBN}$, where $\varepsilon_0$ is the vacuum permittivity. For our device, thickness of the $SiO_2$ bottom dielectric and hBN top dielectric are $d_{SiO2} = 285$ nm and $d_{hBN} = 14$ nm, respectively, and we used a relative dielectric constant of 3.9 for both $SiO_2$ and hBN. Electrical transport measurements were performed in a closed-cycle He cryostat down to 11 K. Standard low-frequency (13.33 Hz) lock-in techniques were employed to measure the device resistance, with 20 mV AC excitation voltage applied together with 100 mV DC bias voltage. The applied DC bias mitigates the effect from the Schottky barrier, allowing measurement of the intrinsic transport behavior of 2D $A_2T_3X_4$. Further details on contact characterization are provided in Supplementary Section 7.

**First-principles calculations.** DFT calculations were performed using the Vienna ab initio simulation package[50,51]. We used both the Perdew-Burke-Ernzerhof functional (PBE)[52] and the Heyd-Scuseria-Ernzerhof hybrid functional (HSE)[53] for the exchange-correlation energy. The pseudopotential was defined based on the projector augmented-wave method[54]. We observed that the HSE functional, which incorporates a fraction of Hartree-Fock exchange energy, is required to accurately describe experimental band structures of $A_2T_3X_4$, similar to the cases of other intermediate bandgap semiconductors. The fraction of Hartree-Fock exchange energy ($a_{HF}$) in HSE functional was determined for each compound in the $A_2T_3X_4$ series by comparing to experimental data. Specifically, we employed $a_{HF}$ = 5 %, 5 %, 15 %, and 15 % for $Cs_2Ni_3S_4$, $Cs_2Ni_3Se_4$, $(Rb,Cs)_2Pd_3S_4$, and $(Rb,Cs)_2Pd_3Se_4$, respectively. An overall renormalization factor was applied to DFT calculation to match experimental dispersion, as discussed in Fig. 2l. All calculations were performed with a plane-wave energy cut-off at 450 eV and a Γ-centered 5×5×1 reciprocal-space mesh for monolayers and 4×4×4 reciprocal-space mesh for bulk $A_2T_3X_4$. To facilitate comparison between bulk and monolayer electronic structure, we plotted bulk band structure using projected Brillouin zone throughout the manuscript (see Extended Data Fig. 8 for details). To understand orbital characters, we performed orbital projections of DFT dispersions onto $T$-$d_{zx}$, $d_{z2-x2}$, $d_{xy}$, $d_{yz}$, $d_{3y2-r2}$ basis. The details of crystal field and $T$-$d$ orbital energy level splitting in $Rb_2Pd_3S_4$ are discussed in Extended Data Fig. 1.

**Tight-binding parameterization.** For the mean-field phase diagram calculations and exact diagonalization analysis, we constructed tight-binding models for $A_2Pd_3S_4$, as shown in Extended Data Fig. 10. Our model is based on a single-orbital, spinful tight-binding Hamiltonian with parameters fitted to the DFT band structure, including hopping terms up to the seventh nearest neighbor, with intrinsic nearest-neighbor spin-orbit coupling.[55] Full details of the tight-binding parameterization and fit to the DFT dispersion are discussed in Supplementary Section 8. We confirm that the mean-field phase diagrams (Fig. 5f) and fractional ground states (Fig. 5i,j) obtained for $A_2Pd_3S_4$ remain robust against the choice of tight-binding parameters and fitting procedures, as also discussed in Supplementary Section 8.

**Mean field phase diagram.** The mean field phase diagram is obtained by solving self-consistent mean-field equations for local magnetic moments and charge densities on a 12×12×3 lattice with periodic boundary conditions[22]. The calculation is based on a tight-binding model with standard onsite Hubbard interactions. The Hubbard interaction is decomposed into local spin and charge channels to construct the Hartree-Fock Hamiltonian[22], which can be solved iteratively. More precisely, we solve for the density matrix variationally at fixed filling using an iterative energy minimization procedure. Starting from an initial density matrix, we construct the corresponding Hartree-Fock Hamiltonian, diagonalize it, and repopulate the lowest-energy single-particle states to obtain an updated density matrix. This process is repeated until both the energy density and the density matrix elements converge within a tolerance of $10^{-14}$ relative to the nearest-neighbor hopping strength. When numerical self-consistency is reached, the values of the order parameters are computed and used to determine the phase.

**Exact Diagonalization Calculations.** The exact diagonalization calculations shown in Fig. 5i, Extended Data Fig. 9, and Supplementary Section 8 were performed on the topmost flat Chern band of $A_2Pd_3S_4$, that emerges within the ferromagnetic phase appearing over a broad range of parameter space as shown in Fig. 5f. The topmost flat Chern band carries the Chern number 1, mimicking the lowest Landau level. The Hamiltonian is constructed as the sum of the band dispersion and the band-projected Coulomb interaction. To ensure numerical convergence of the momentum-space summation in the interaction term, we use the screened Coulomb potential $V(q) \sim e^{-qd}$ and decrease $d$ until the results become insensitive to further reductions. We assume a dielectric constant $\varepsilon_r = 5$ corresponding to hexagonal boron nitride encapsulation. Band mixing between the topmost flat Chern band and the second-highest band is incorporated by allowing at most one electron to be excited from the lower band into the flat band[56]. This captures the leading mixing processes while keeping the Hilbert space manageable, enabling exact diagonalization calculations on sufficiently large momentum grids. Further details are provided in Supplementary Section 8. Sparse diagonalization is performed using Krylov methods. For each total momentum $K$ sector, the energy spectrum includes up to the six lowest eigenvalues. By adopting a gauge-invariant discretized Berry-phase method[57], we compute the many-body Chern number of the ground states over twisted boundary conditions, which is obtained as $C = 1/3$, consistent with the spectral flow (Fig. 5j).

**Acknowledgements**

We acknowledge Advanced Light Source, Diamond Light Source, and MAX-IV Laboratory for provision of nano- and micro-ARPES beamtime under Proposal #12800 (ALS), #SI38510-1 (DLS), and #20231702 (MAX-IV). This work was primarily supported by the National Science Foundation through Grant No. DMR-2104427 (P.M., C.M., M.K. K.M.S.), the Air Force Office of Scientific Research through Grant No. FA955021-1-0168 and Grant No. FA9550-23-1-0161 (P.M., C.M., M.K., K.M.S.), and the National Research Foundation (NRF) of Korea through Grant No. RS-2025-23525391, RS-2026-25496507 (H.B., Sehyun Oh, J.H., H.L., J.C., M.K.) and Grant No. RS-2022-NR068223, RS-2026-25585727 (S.L., C.W., J.K., J.-H.P.). M.K. was supported by Creative-Pioneering Researchers Program through Seoul National University. Soomin Oh and G.Y.C. are supported by Samsung Science and Technology Foundation under Project Number SSTF-BA2401-03, the NRF of Korea (Grants No. RS-2026-25479545, RS-2024-00410027, RS-2024-00444725, RS-2023-00256050, RS-2025-25453111, RS-2025-08542968) funded by the Korean Government (MSIT), the Air Force Office of Scientific Research under Award No. FA23862514026, and Institute of Basic Science under project code IBS-R014-D1. Soomin Oh acknowledges the support from Agency for Defense Development through a grant funded by the Defense Acquisition Program Administration (DAPA) (UI257011TE). This study was in part funded by the Korea Basic Science Institute (S.-H.R., grant number C526110). This research used resources of the Advanced Light Source, which is a DOE Office of Science User Facility under contract no. DE-AC02-05CH11231. H.B. acknowledges supports from Hyunsong Educational & Cultural Foundation. J.K. acknowledges a PIURI postdoctoral fellowship from POSTECH. M.K. acknowledges a Kavli Postdoctoral Fellowship from the Kavli Institute at Cornell and POSCO Science Fellowship from Cheongam Foundation.

**Author contributions**

M.K. and K.M.S., G.Y.C. and J.-H.P. conceived the project; C.W. synthesized the bulk crystals; H.B., P.X.N., R.C., H.L., and M.K. fabricated 2D devices under the supervision of J.L., J.S., and K.F.M.; H.B., S.L., Sehyun Oh, J.H., P.M., C.M., J.K., S.-H.R., and M.K. performed the ARPES experiments with support from C.J., A.B., E.R., C.M.P., M.D.W., and T.K.K.; S.L. carried out the DFT calculation; Soomin Oh and G.Y.C. performed the mean-field calculations and exact diagonalization analysis; S.Y and H.Y. supported the AFM measurements; J.C. assisted with ARPES data analysis; H.B., Soomin Oh, J.-H.P., K.M.S., G.Y.C., and M.K. wrote the manuscript with inputs from all co-authors.

## Data availability

The datasets presented within this study are available from the corresponding authors upon reasonable request.

## Competing interests

The authors declare no competing interests.

## Methods Only References

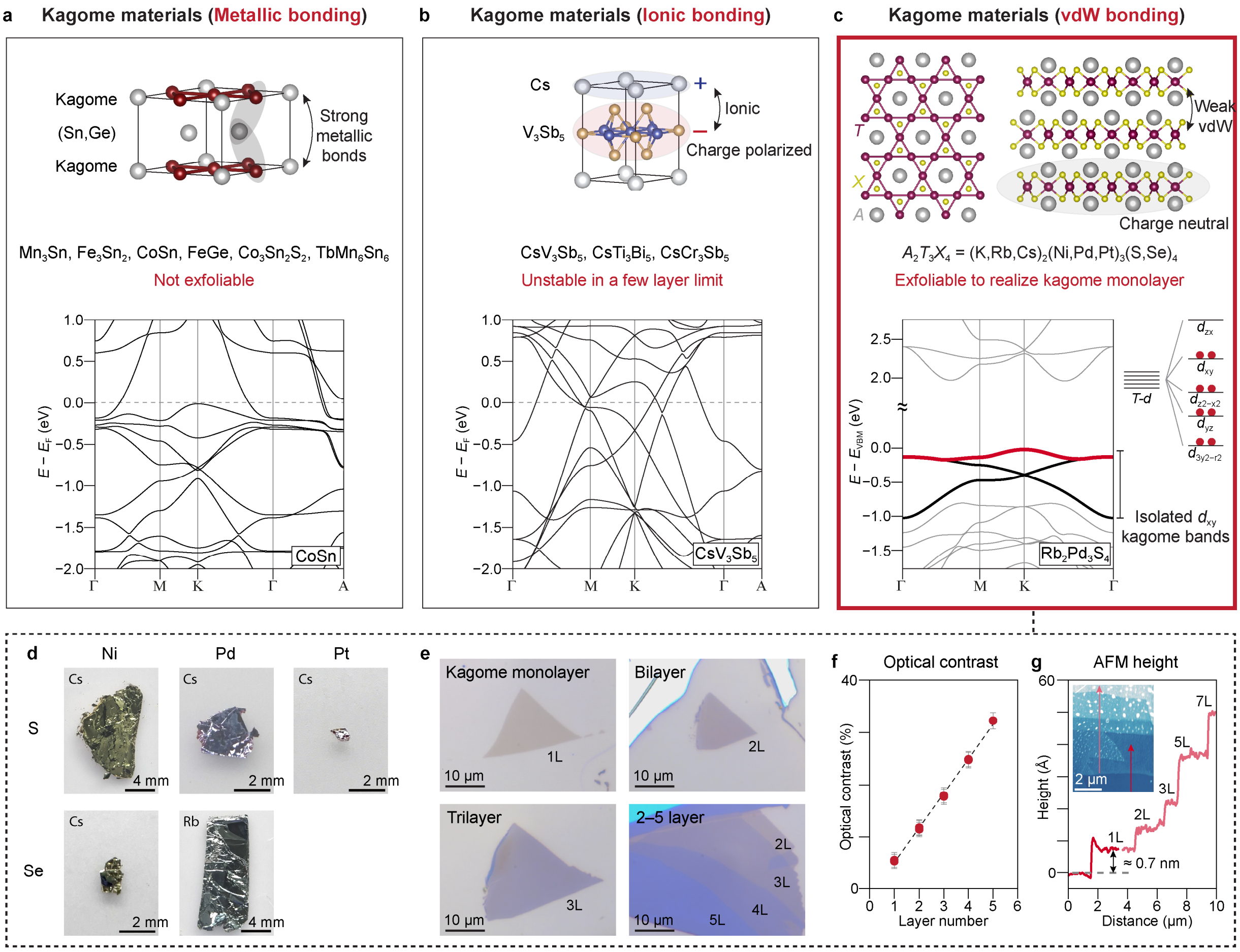

**Figure 1 | vdW kagome materials family $A_2T_3X_4$ as a new class of 2D flat band system. a–c,** Classification of kagome lattice materials based on type of interlayer bonding: metallic (a), ionic (b), and vdW (c). The vdW kagome family $A_2T_3X_4$ not only offers an ideal opportunity to realize true 2D monolayer kagome lattices, but also hosts an isolated kagome flat band at the valence band maximum ($E_{VBM}$). **d,** Bulk crystals of $A_2T_3X_4$. **e,** Realization of the 2D kagome lattices by mechanical exfoliation of $Cs_2Pd_3S_4$ down to the monolayer limit. **f,g,** Optical contrast and atomic force microscopy measurements on exfoliated $Cs_2Pd_3S_4$, confirming single-layer thickness. The inset in g displays an atomic force microscopy image from which the height profiles are extracted (red and coral solid arrows).

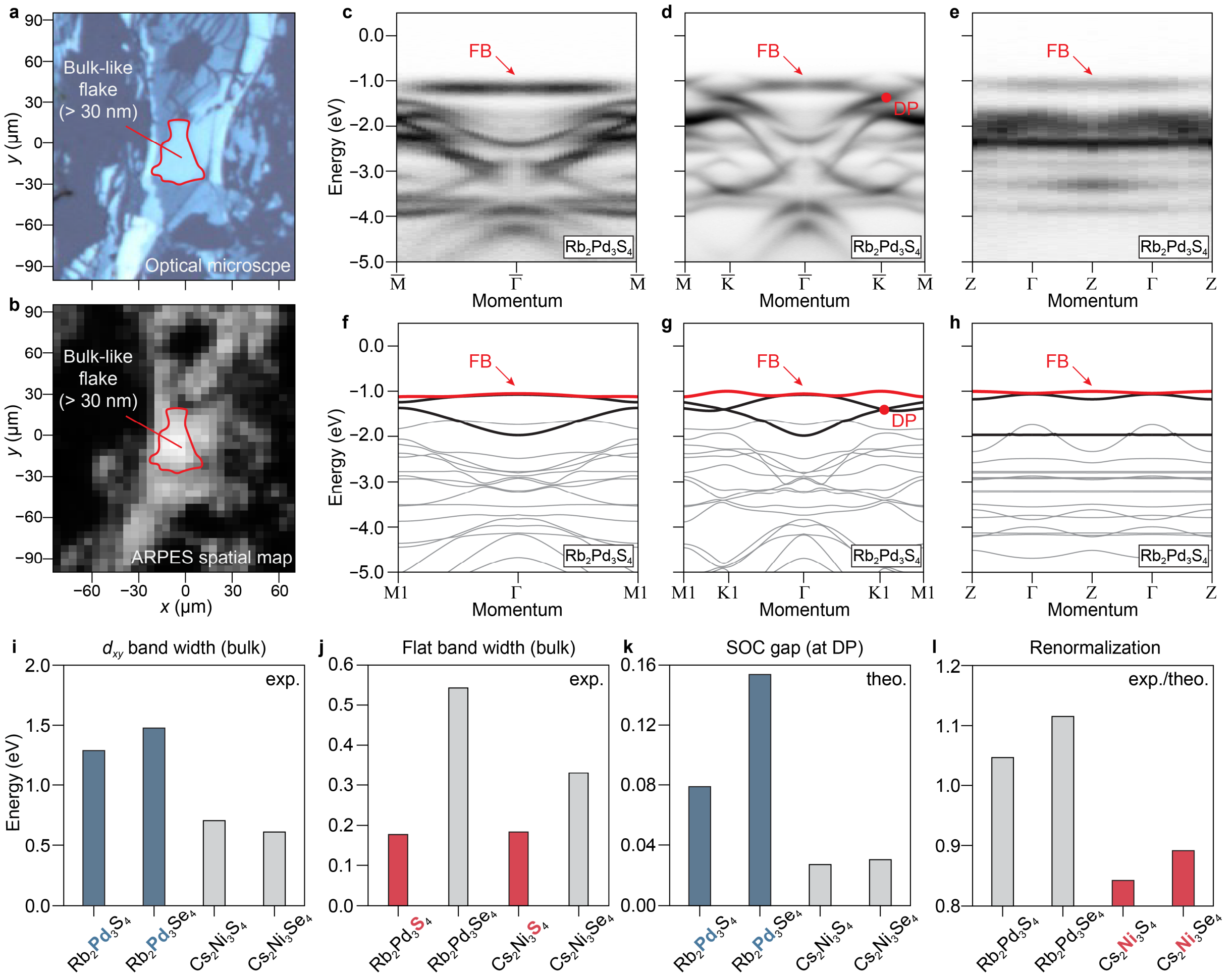

**Figure 2 | Bulk electronic structure of the vdW kagome series $A_2T_3X_4$. a,** Optical microscope image of exfoliated $A_2T_3X_4$ flakes on conducting substrate. Among the exfoliated flakes, we selected a thick-enough flake (> 30 nm) to obtain bulk electronic structure of $A_2T_3X_4$. Compared to conventional ARPES measurements on cleaved bulk crystals, this approach reduces sample charging, yielding substantially higher ARPES spectrum quality. **b,** Corresponding ARPES spatial map visualizing the exfoliated flakes *in situ*. **c,d,** ARPES spectra of $Rb_2Pd_3S_4$ measured along $\overline{\Gamma}$–$\overline{M}$ and $\overline{\Gamma}$–$\overline{K}$ high-symmetry surface Brillouin zone directions, respectively. For comparison, the corresponding DFT band structures along Γ–M1 and Γ–K1 are shown in **f,g** (see Extended Data Fig. 8 for the definition of projected Brillouin zone of $A_2T_3X_4$). Red and black solid lines highlight the $d_{xy}$ kagome bands, forming an isolated quasi-flat band and Dirac bands near the Fermi level. **e,h,** Out-of-plane dispersion of $Rb_2Pd_3S_4$ obtained from photon-energy-dependent ARPES measurements (e) and DFT (h). ARPES spectra in c–e were symmetrized with respect to the center of the image. **i–l,** Systematic evolution of band parameters across the $A_2T_3X_4$ kagome family – $d_{xy}$ band width (i), flat band width (j), spin-orbit coupling gap at the Dirac point (k), and the degree of electronic correlation (band renormalization factor) (l). The bandwidths (i,j) were extracted from the experimental ARPES spectra, while SOC gap (k) were obtained from DFT dispersions.

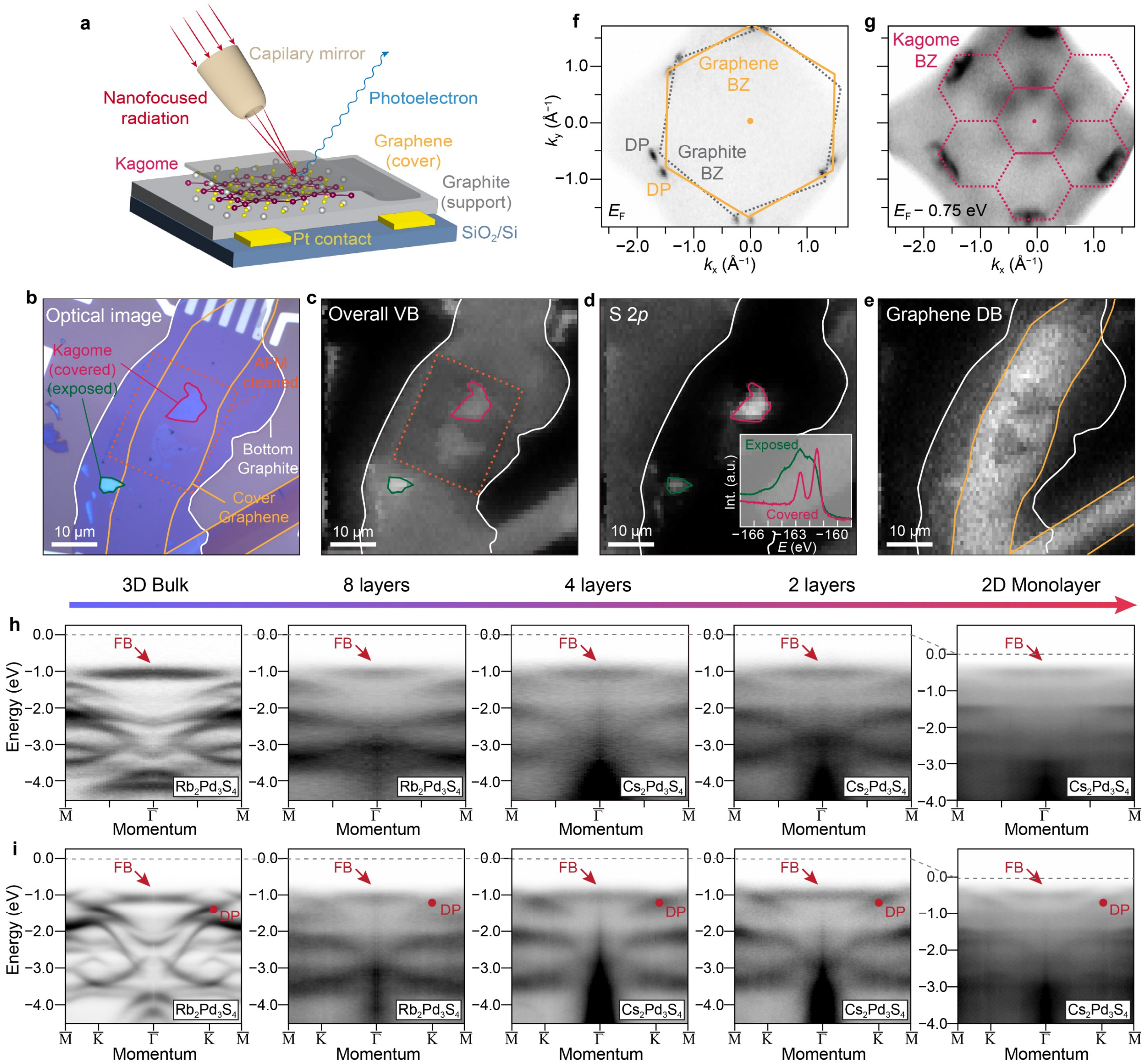

**Figure 3 | Electronic structure of 2D $A_2Pd_3S_4$ measured by nano-ARPES. a,** Schematics of the graphene/$A_2Pd_3S_4$/graphite vdW heterotrilayer prepared for nano-ARPES measurements. **b,** Optical image of a representative heterotrilayer device. **c–e,** Spatial mappings of photoelectron intensities using the valence band photocurrents (c), S-2*p* core levels (d), and graphene Dirac bands (e), respectively. These mappings offer *in situ* visualization of bottom graphite, $A_2Pd_3S_4$ flakes, and top graphene, respectively, as marked with white, magenta, and yellow solid lines. Green solid lines mark an additional $A_2Pd_3S_4$ flake on our device not covered by graphene, while orange-dashed lines indicate a surface region cleaned by an atomic force microscopy tip (see Methods). The inset in d compares S-2*p* core level spectra of graphene-covered versus exposed $A_2Pd_3S_4$ flakes. **f,g,** Fermi surface (f) and constant energy contour at 0.75 eV binding energy (g) of the graphene/$A_2Pd_3S_4$/graphite heterostructure. Overlaid yellow, gray, and magenta hexagons mark the Brillouin zones of graphene, graphite, and $A_2Pd_3S_4$, respectively. **h,i,** Evolution of electronic structure from 3D to 2D $A_2Pd_3S_4$. The flat band and Dirac point clearly persist down to the monolayer limit of $A_2Pd_3S_4$. ARPES spectra were symmetrized with respect to the center of the image.

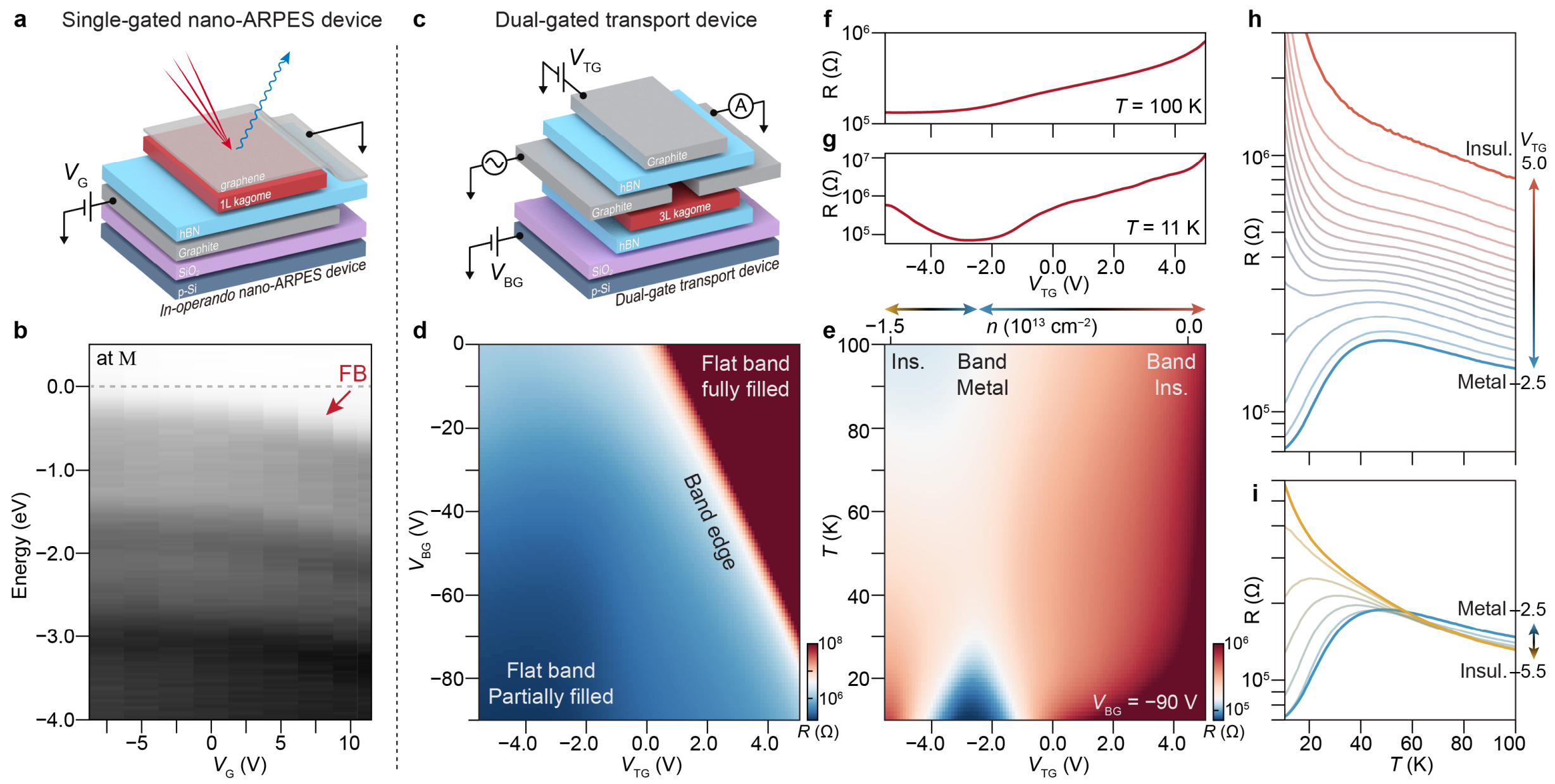

**Figure 4 | Gate-engineering and access to the partially filled kagome flat band states in 2D $A_2T_3X_4$. a,** Schematic illustration of the single-gated $A_2T_3X_4$ field effect device used for *in operando* nano-ARPES measurements. The device is composed of [cover graphene / monolayer $Cs_2Pd_3S_4$ / hBN/graphite bottom gate]. **b,** Gate-dependent evolution of energy distribution curves measured at the M point, spectroscopically visualizing the gate-tunability of the kagome flat band. **c,** Schematic illustration of the dual-gated $A_2T_3X_4$ field effect device used to measure two-terminal transport. The device consists of [graphite/hBN top gate / graphite electrodes / trilayer $Rb_2Pd_3Se_4$ / $SiO_2$/Si bottom gate]. **d,** Resistance map as a function of top- and bottom-gate voltages measured at 70 K. Clear gate-induced resistivity reduction across the band edge is observed, demonstrating successful carrier-doping and experimental access to the partially filled kagome flat band regime. **e,h,i,** Temperature-dependence of the resistance measured at various $V_{TG}$, with the bottom gate fixed at $V_{BG} = -90$ V. Corresponding carrier density in e is estimated based on a simple parallel plate capacitor model (see Methods). In the neutral or weakly hole-doped regime (0V < $V_{TG}$ < 5 V), the device displays insulating temperature dependence (reflecting band insulator or disorder-localized insulator states), whereas metallic transport from partially filled flat band electrons is directly observed upon further hole-doping (–2.5 V < $V_{TG}$ < 0 V, e,h). Beyond this band metal regime, another insulating state appears within the partially filled kagome flat band (–5.5 V < $V_{TG}$ < –2.5 V, e,i). **f,g,** Line cuts of e at $T$ = 100 K and 11 K, respectively. The monotonic decrease of the resistance with increasing negative $V_{TG}$ at 100 K confirms continuous hole-doping to the flat band states (f), while the increase of the resistance beyond $V_{TG}$ < –2.5 V at 11 K signifies the manifestation of the insulating regime (g).

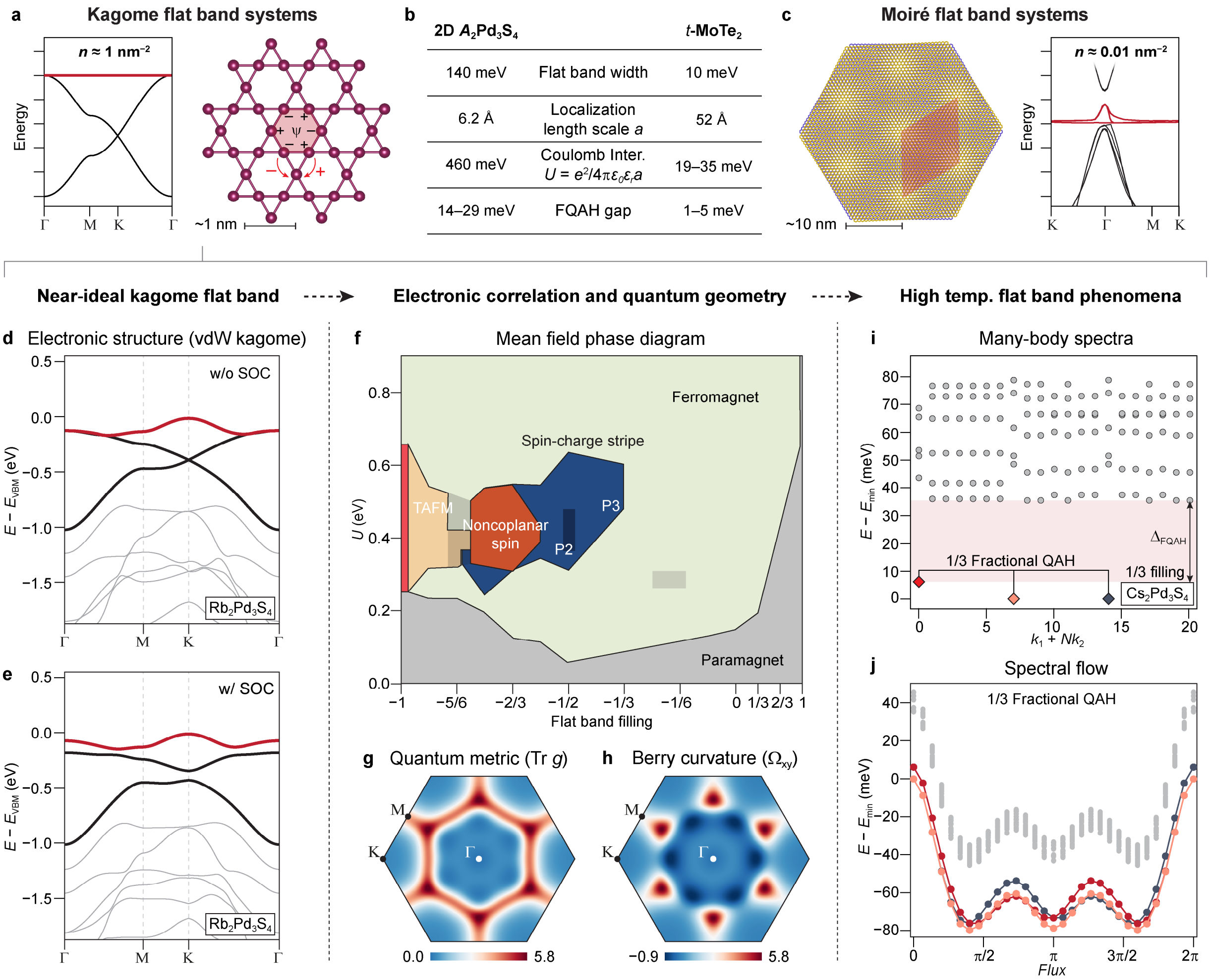

**Figure 5 | Potential for high-temperature flat band phenomena in 2D $A_2T_3X_4$. a–c,** Comparison between characteristic energy scales in 2D vdW kagome system ($A_2Pd_3S_4$) and Moiré superlattice system (twisted bilayer $MoTe_2$). The latter is adopted from Ref. 42. Flat bands in kagome lattice hosts two orders of magnitude higher flat band electron density ($n$) compared to Moiré systems and, consequently, enhanced energy and temperature scales. **d,e,** Electronic structure of monolayer $Rb_2Pd_3S_4$ without (d) and with (e) spin-orbit coupling. **f,** Mean-field phase diagram of monolayer $Rb_2Pd_3S_4$ at partial fillings of the flat band. A cascade of strongly correlated phases emerges in different parameter regimes, including 120° triangle antiferromagnetism (TAFM), topological noncoplanar spin states with nonzero scalar spin chirality, spin-charge stripe phases with various periodicities (P2, P3), and flat band ferromagnetism, which is the well-known exact ground state of the kagome flat band at half-filling[20]. **g,h,** Quantum metric ($g_{xx}+g_{yy}$) and Berry curvature ($\Omega_{xy}$) of the flat band in $Rb_2Pd_3S_4$. **i,** Exact many-body spectrum of the fractional-filled topmost Chern band in $Cs_2Pd_3S_4$. Corresponding spectrum for $Rb_2Pd_3S_4$ is presented in Extended Data Fig. 9. The characteristic threefold ground-state quasi-degeneracy indicates a Laughlin-type state analogous to the fractional-filled lowest Landau level, supporting the stabilization of the fractional quantum anomalous Hall state (FQAH). The FQAH state in 2D $A_2Pd_3S_4$ is characterized by a large excitation gap ($\Delta_{FQAH} \approx$ 14–29 meV), as marked by red-shaded area. **j,** Spectral flow of the many-body spectra under the insertion of one flux quantum. The quasi-degenerate ground states undergo interchange, and return to the original configuration only after the 6π flux insertion, consistent with fractional ground states characterized by many-body Chern number 1/3.

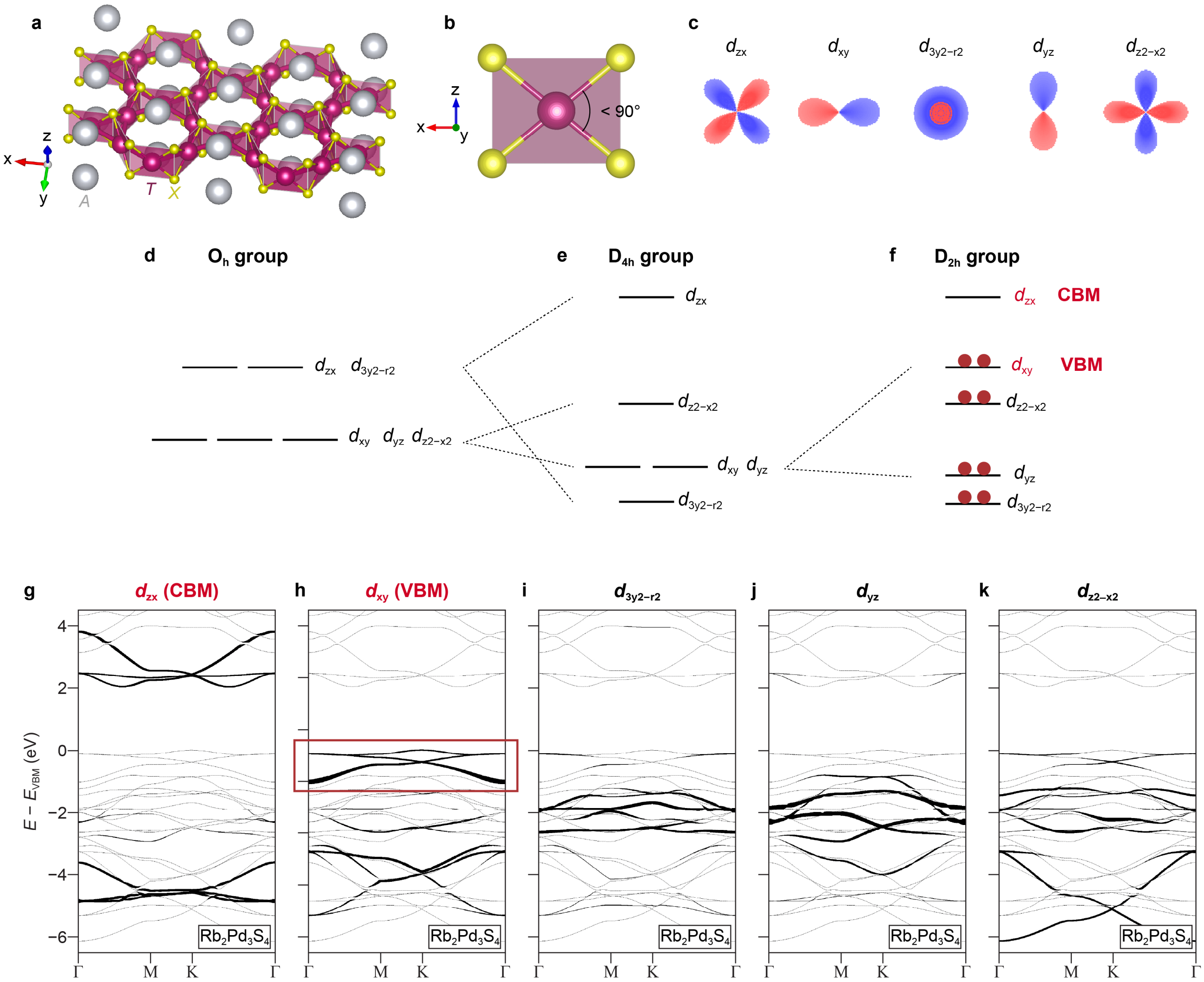

**Extended Data Figure 1 | Crystal field splitting and isolation of the $d_{xy}$ kagome bands in $A_2T_3X_4$. a,b,** Lattice structure (a) and local ligand environment (b) of the $A_2T_3X_4$ vdW layer. The kagome network of $A_2T_3X_4$ is constructed by transition metal $T$ atoms as shown in a. Each $T$ site is coordinated by four chalcogen ligands $X$, forming local $D_{2h}$ symmetry as highlighted in b. Throughout the manuscript, we define a local Cartesian coordinate at each $T$ site, with $z$-axis normal to the kagome-plane, $y$-axis normal to the ligand plane, and $x$-axis orthogonal to both. This convention defines the kagome plane as the $xy$ plane and the ligand plane as the $zx$ plane. **c,** Schematic illustration of corresponding $d$-orbitals ($d_{zx}$, $d_{z2-x2}$, $d_{xy}$, $d_{yz}$, $d_{3y2-r2}$) at the $T$ site. **g–k,** Orbital-projected band structure of monolayer $A_2T_3X_4$. We present the representative case of $Rb_2Pd_3S_4$ here, but all $A_2T_3X_4$ exhibit similar orbital hierarchy. The conduction band minimum (CBM) and valence band maximum (VBM) are primarily composed of $d_{zx}$ and $d_{xy}$ orbitals, respectively, while other $d$ orbitals are pushed well below the Fermi level. The isolated $d_{xy}$ kagome bands, which is the main focus of the present manuscript, is highlighted in h. The $d_{xy}$ orbital character of the VBM bands in $A_2T_3X_4$ is further supported by polarization-dependent ARPES measurements, as discussed in detail in Supplementary Section 5. One can intuitively understand this orbital level splitting in $A_2T_3X_4$ by considering strong $D_{2h}$ crystal field, as schematically illustrated in d–f.

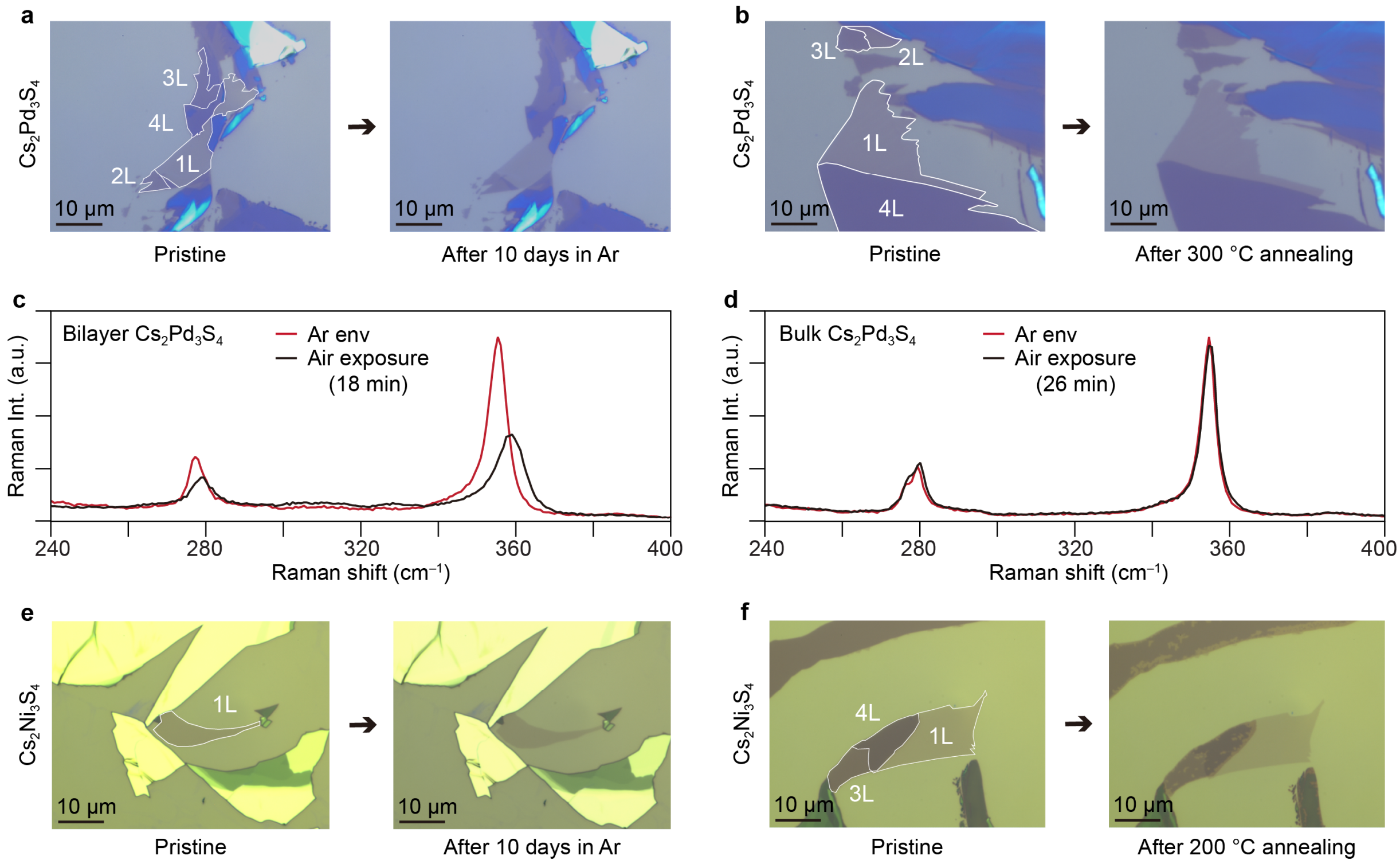

**Extended Data Figure 2 | Stability of exfoliated 2D $A_2T_3X_4$. a,** Optical microscope images of exfoliated $Cs_2Pd_3S_4$ flakes taken immediately after exfoliation (left) and after 10 days in an Ar glovebox (right). No degradation is observed under optical inspection. **b,** Thermal stability of exfoliated $Cs_2Pd_3S_4$ flakes. The flakes remain stable under optical inspection up to 300 °C. **c,d,** Characterization of air-sensitivity of exfoliated $Cs_2Pd_3S_4$ flakes using Raman spectroscopy. Under an Ar atmosphere, the $Cs_2Pd_3S_4$ flakes exhibit sharp Raman peaks around 280 cm$^{-1}$ and 355 cm$^{-1}$, corresponding to S vibrations. Upon exposure to air for 10–30 min, a bulk-like $Cs_2Pd_3S_4$ flake with thickness > 20 nm shows minimal change in the Raman spectra, indicating marginal air-sensitivity (d). In contrast, bilayer $Cs_2Pd_3S_4$ exhibits clear frequency shifts and spectral broadening, indicating enhanced air-sensitivity upon approaching the 2D limit (c). We note that such increased air sensitivity in the 2D limit is typical for vdW materials, as previously observed in black phosphorus, $CrI_3$, $MoTe_2$, $NiPS_3$, etc. **e,f,** Corresponding characterizations for $Cs_2Ni_3S_4$. In contrast to the case of $Cs_2Pd_3S_4$ in a,b, substantial degradation is optically apparent already at 200 °C, indicating that environmental stability depends on composition within the $A_2T_3X_4$ family. Regarding the stability of 2D $A_2T_3X_4$, we also refer to Ref. 35, in which the thermodynamical and mechanical stability of monolayer $A_2T_3X_4$ was theoretically predicted.

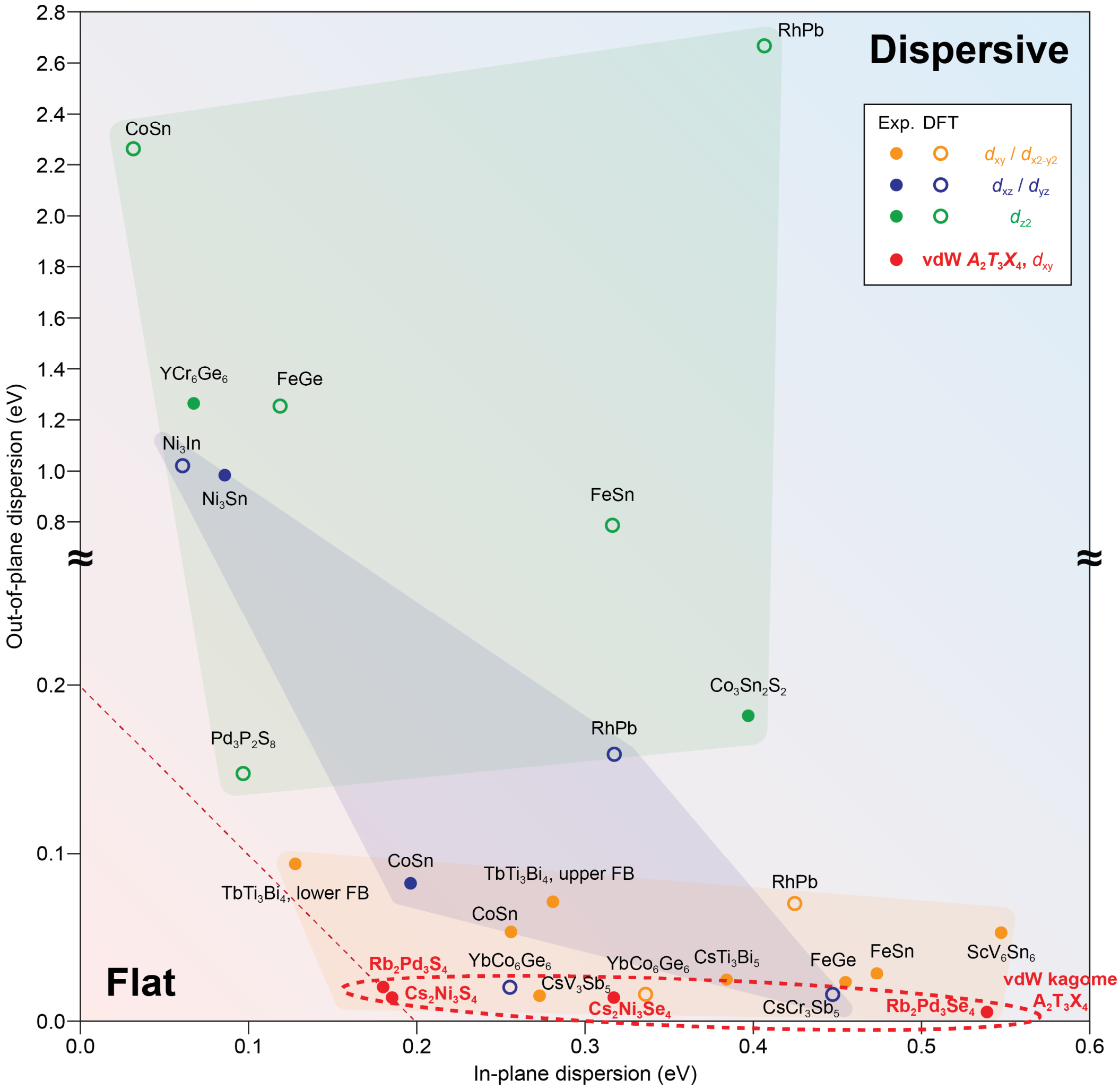

**Extended Data Figure 3 | Survey of flat bands in kagome systems.** Unlike the idealized kagome model, flat bands in real kagome materials acquire finite bandwidth due to additional factors neglected in the toy model. Some of the important factors affecting flat band dispersion include beyond nearest-neighbor hopping, inter-kagome-layer interactions, multiple orbital degrees of freedom, etc. The diagram above summarizes flat band dispersions across various kagome lattice materials reported to date, with the *x*- and *y*-axes denoting in-plane and out-of-plane dispersions, respectively. Note that we only considered the cases where the flat band lies within 1 eV of the Fermi level. Despite the complex factors affecting flat band bandwidth, several clear trends can be observed in the diagram. First, flat bands with in-plane orbital characters ($d_{xy}/d_{x2-y2}$, orange) exhibit weak out-of-plane dispersion (< 0.1 eV) while tend to develop pronounced in-plane bandwidth, likely due to enhanced longer-range hopping within the kagome plane. By contrast, flat bands with out-of-plane orbital characters ($d_{z2}$, green) exhibit opposite behavior, featuring significant out-of-plane dispersion (reaching several eV) but relatively small in-plane bandwidth. Flat bands with $d_{xz}/d_{yz}$ orbital character (blue) fall between these two limits. The vdW kagome family $A_2T_3X_4$ populates the bottom of the diagram, characterized by negligible out-of-plane dispersion owing to weak vdW interlayer coupling. Notably, taking into account both in-plane and out-of-plane dispersions, the flat bands in $Rb_2Pd_3S_4$ and $Cs_2Ni_3S_4$ realize among the flattest dispersion in kagome lattice materials reported to date.

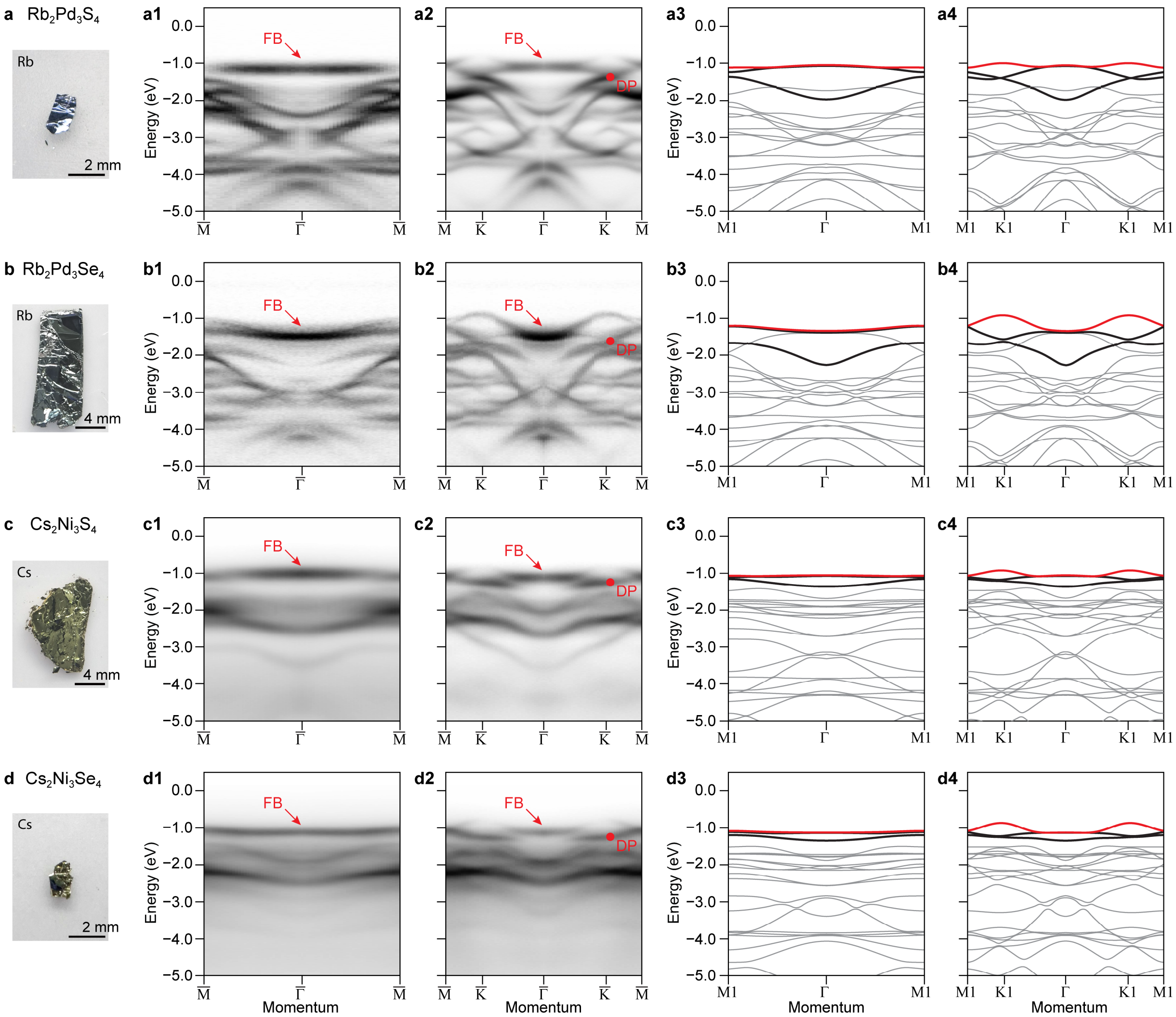

**Extended Data Figure 4 | Systematic catalogue of electronic structures in the $A_2T_3X_4$ vdW kagome family.** **a–d,** ARPES spectra of $Rb_2Pd_3S_4$, $Rb_2Pd_3Se_4$, $Cs_2Ni_3S_4$, and $Cs_2Ni_3Se_4$ measured along the $\bar{\Gamma}$–$\bar{M}$ (a1–d1) and $\bar{\Gamma}$–$\bar{K}$ (a2–d2) high symmetry directions of the surface Brillouin zone. The spectra were symmetrized with respect to $\bar{\Gamma}$. The corresponding DFT band structures are also displayed in a3–d3 and a4–d4, respectively. As highlighted by red and black solid lines, all compounds in the $A_2T_3X_4$ series exhibit an isolated quasi-flat band and Dirac bands just below the Fermi level. Despite this overall universality, the detailed band parameters – such as the bandwidth of the $d_{xy}$ manifold, flatness of the quasi-flat band, spin-orbit coupling gap at the Dirac point, and the degree of electronic correlation (band renormalization factor) – are highly tunable within this $A_2T_3X_4$ kagome family, as discussed in Fig. 2i–l.

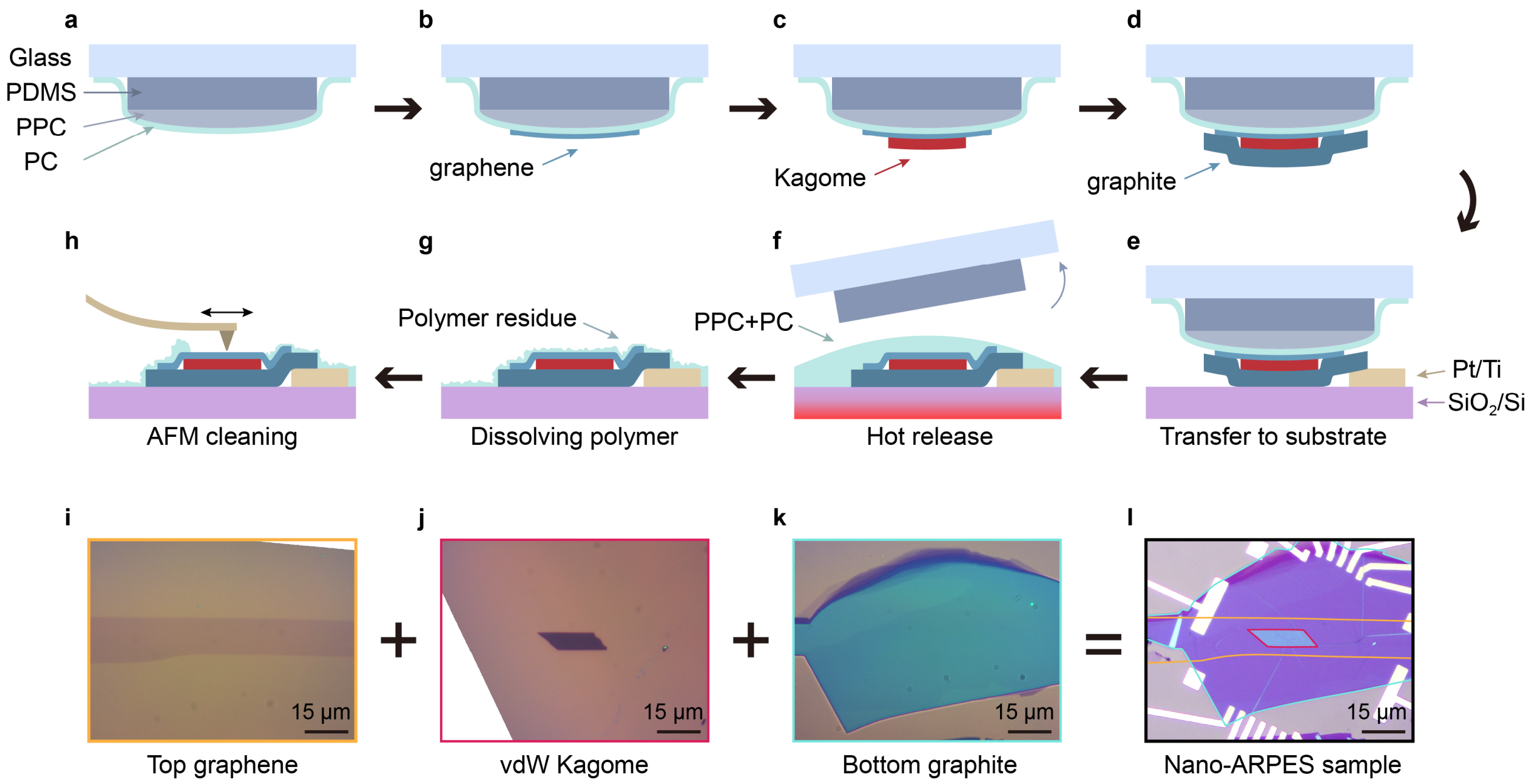

**Extended Data Figure 5 | Fabrication of the 2D kagome-based vdW heterostructure for nano-ARPES study. a,** Structure of the dome-shaped PC/PPC/PDMS stamp. **b–d,** Sequential pick-up of graphene (b), kagome (c), and graphite (d) flakes to assemble the heterotrilayer stack. The 2D kagome flake is encapsulated between graphene and graphite, being protected against air and solvent exposure in the subsequent steps. **e,** Transfer of the stack onto the substrate, with the bottom graphite making contact with the prepatterned Pt/Ti electrode to establish a conduction channel to the kagome flake. **f,** Hot release of the stack by heating above the glass transition temperature of PC, allowing polymers to melt and detach from the PDMS. **g,** Chloroform cleaning for 30 minutes to dissolve PC and PPC. **h,** Mechanical removal of polymer residues from the stack surface using an atomic force microscopy tip. **i–l,** Representative optical images of the top graphene (i), few-layer $Rb_2Pd_3S_4$ (j), bottom graphite (k), and the fully assembled stack (l).

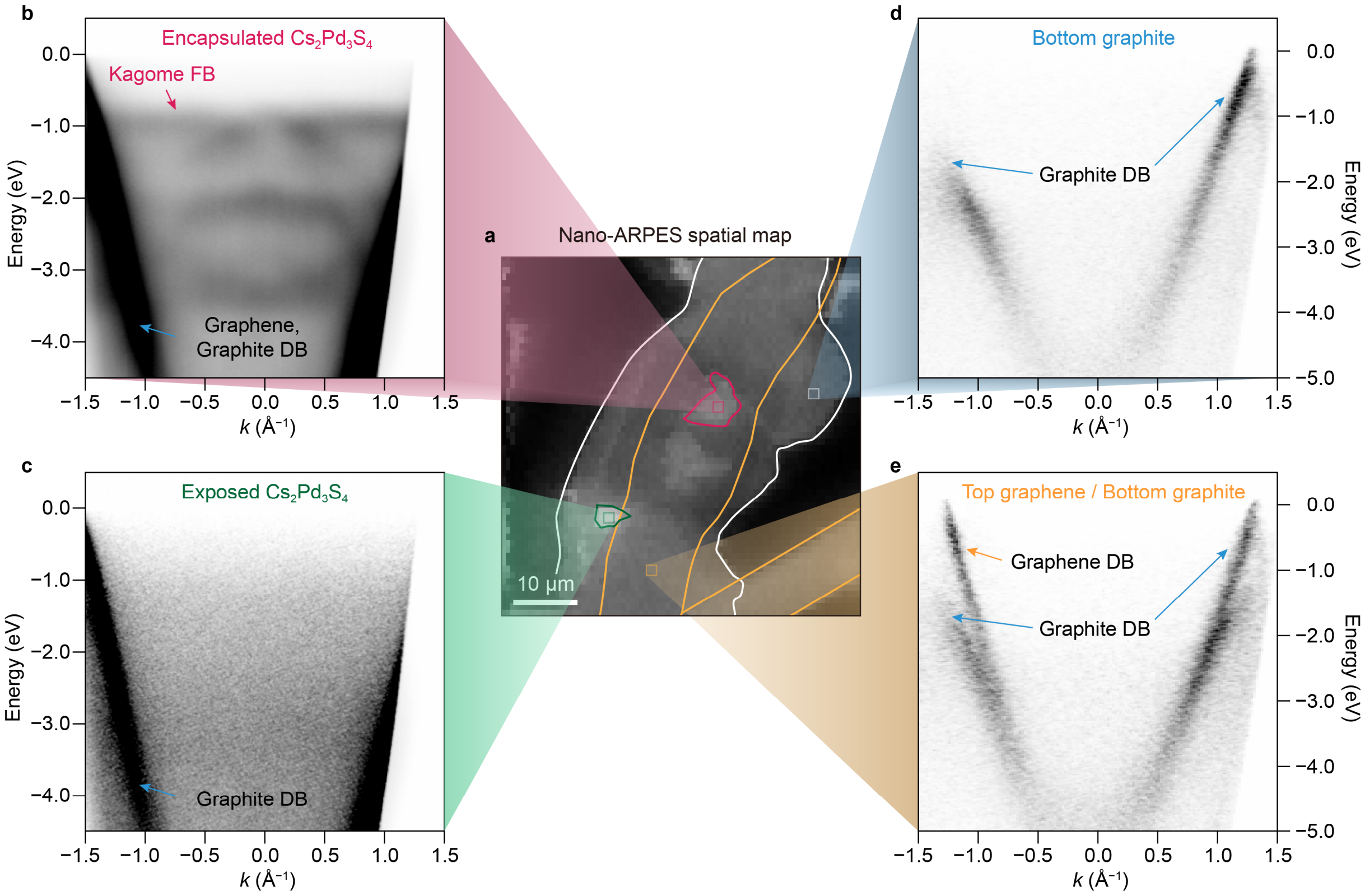

**Extended Data Figure 6 | Spatially-resolved electronic structure of the 2D kagome-based vdW heterostructure. a,** Nano-ARPES spatial map of the [graphene / $Cs_2Pd_3S_4$ / graphite] heterotrilayer, reproduced from Fig. 3c. Yellow and white solid lines mark the top graphene and bottom graphite layers, while magenta and green solid lines indicate the encapsulated and exposed kagome flakes, respectively. **b,** Nano-ARPES dispersion obtained on the encapsulated kagome flakes. Linearly dispersing Dirac bands (DBs) from the top graphene and bottom graphite layers are observed at large momentum ($k$), while multiple kagome-derived bands appear at relatively small $k$. The flat band dispersion at the valence band top of 2D $Cs_2Pd_3S_4$ is apparent. **c,** Nano-ARPES dispersion from the exposed kagome flakes. Kagome dispersions are notably absent, consistent with degradation of the flake indicated by the core-level spectra (Fig. 3d). **d,** A single Dirac band (blue arrow) observed on the bottom graphite region. **e,** An additional Dirac band (yellow arrow) appearing on the region covered by top graphene.

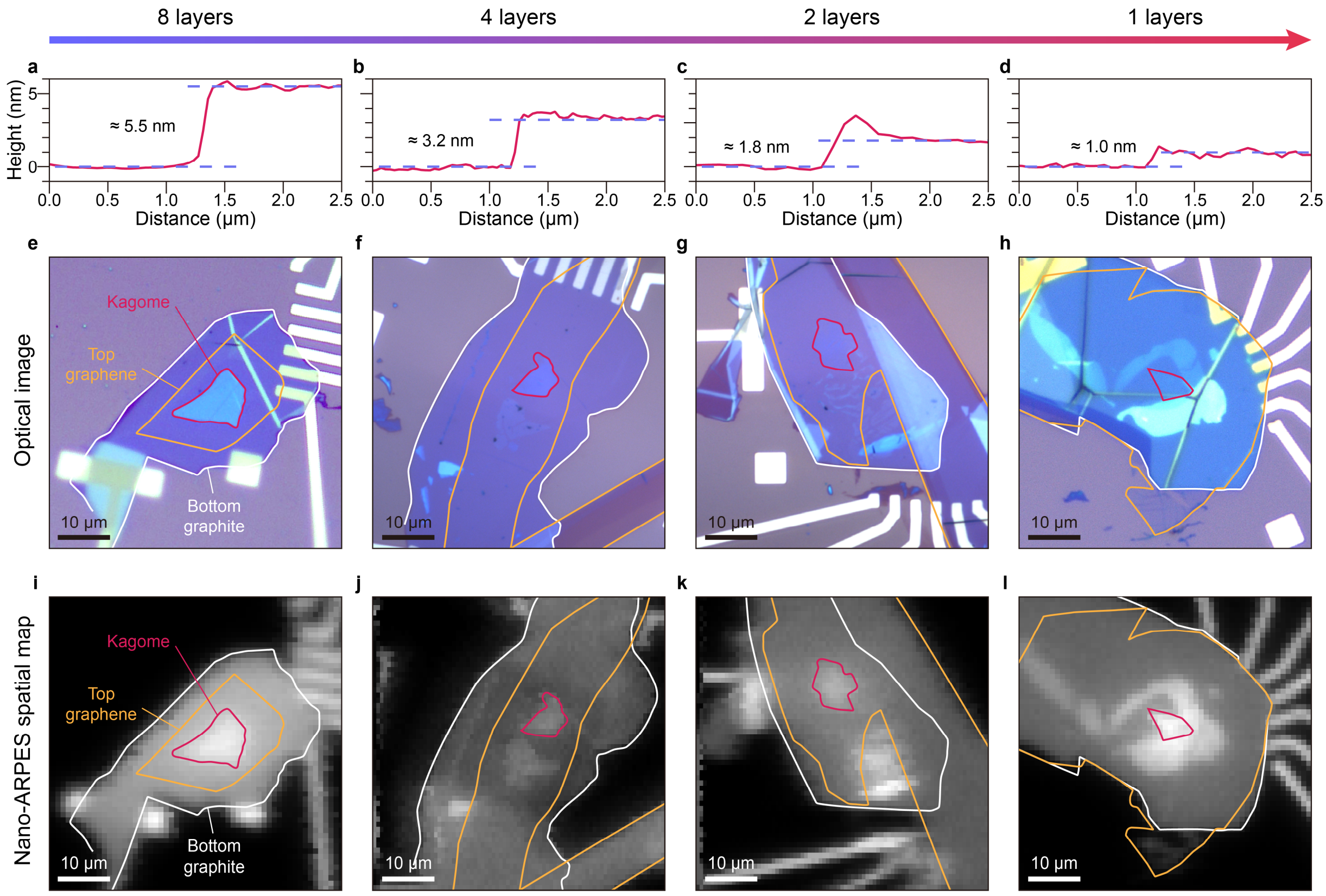

**Extended Data Figure 7 | Nano-ARPES samples with varying thickness of $A_2Pd_3S_4$. a–d,** Atomic force microscopy height profiles measuring 8-, 4-, 2-, and 1-layer $A_2Pd_3S_4$ flakes. The 8-layer flake was $Rb_2Pd_3S_4$ and the 4-, 2-, and 1-layer flakes were $Cs_2Pd_3S_4$. The measured thicknesses are roughly consistent with the interlayer distance of ≈ 0.7 nm. **e–h,** Optical images of [graphene / $A_2Pd_3S_4$ / graphite] vdW devices fabricated using these flakes. Top graphene, bottom graphite, and kagome flakes are marked with yellow, white, and magenta solid lines, respectively. **i–l,** Nano-ARPES spatial maps of the heterostructures, visualizing each layer *in situ*. Corresponding energy-momentum dispersions measured on these nano-ARPES samples are shown in Fig. 3h,i.

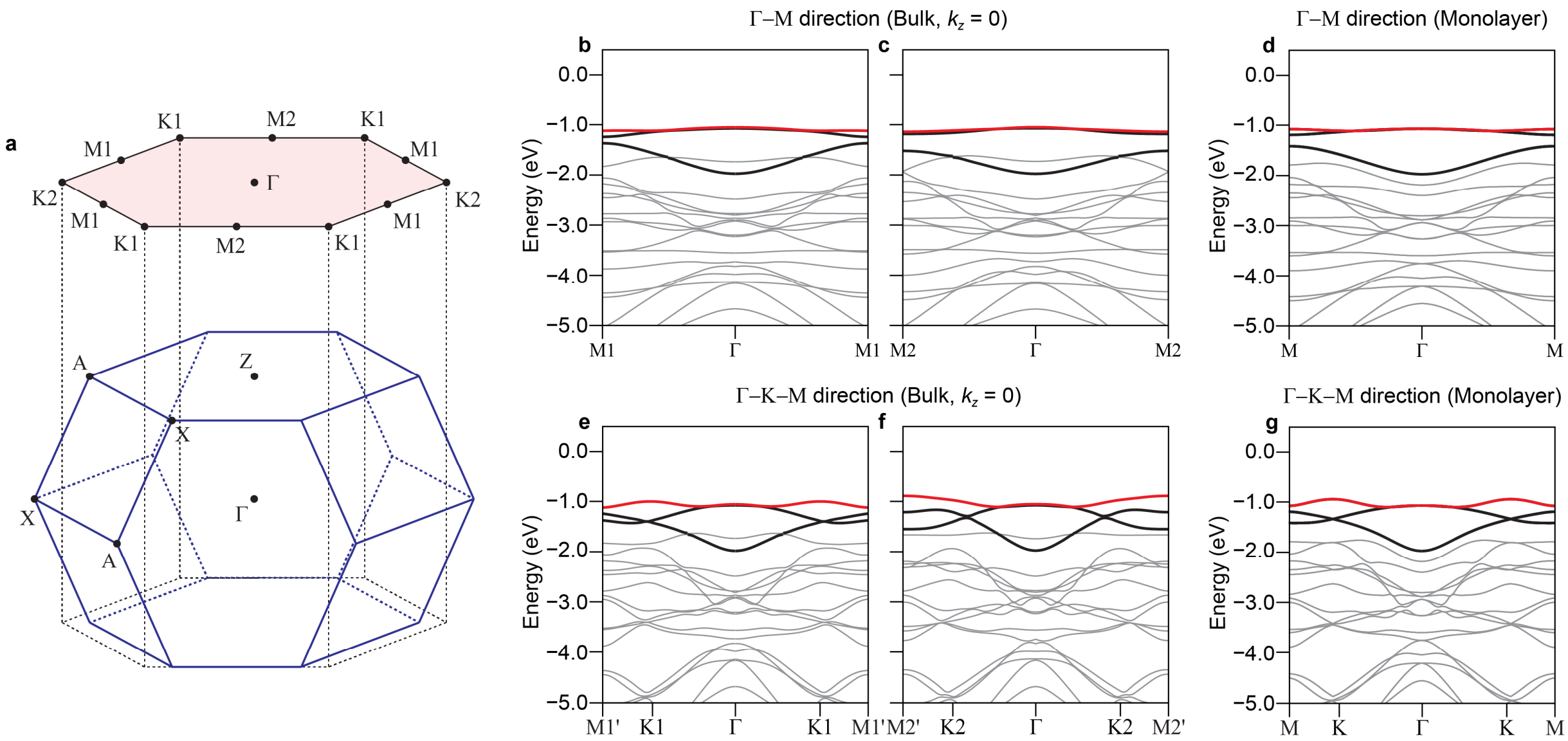

**Extended Data Figure 8 | Bulk versus monolayer electronic structure of $A_2T_3X_4$.** In this Extended Data Figure, we discuss to which degree the kagome electronic structure in 2D monolayer $A_2T_3X_4$ changes in 3D bulk *Fmmm* $A_2T_3X_4$. The conventional unit cell of *Fmmm* is composed of two kagome layers with relative lateral displacements. An important point to note is that the primitive unit cell of *Fmmm* still contains three *T* sites (identical to the number of *T* atoms in a single kagome layer), so that the number of kagome bands does not actually change from 2D to 3D. The bulk Brillouin zone of *Fmmm* $Rb_2Pd_3S_4$ is shown in panel a, along with the projected Brillouin zone used to facilitate comparison with the monolayer $Rb_2Pd_3S_4$ (Scheme similar to used for other *Fmmm* kagome systems, such as $TbTi_3Bi_4$). Panels b–f display the corresponding DFT electronic structures along high-symmetry directions. The most pronounced consequence of the *Fmmm* stacking is breaking of the $C_6$ rotational symmetry, giving rise to the two inequivalent K (M) points in the projected Brillouin zone – labeled as K1 and K2 (M1 and M2), respectively – and asymmetric dispersion along these two directions. Despite these perturbations from the interlayer interaction, the characteristic kagome dispersion of monolayer $Rb_2Pd_3S_4$, i.e. flat band and Dirac bands (panels d,g), clearly persist in the bulk $Rb_2Pd_3S_4$. We note that experimentally, we did not observe pronounced anisotropy along K1 and K2 (M1 and M2) directions in both bulk and few layers devices, suggesting that this effect may be overestimated in the calculations. Accordingly, in Fig. 2 and Extended Data Fig. 4, we present the bulk dispersion along Γ–K1 and Γ–M1 path, which provide closer agreement with the experimental ARPES spectra.

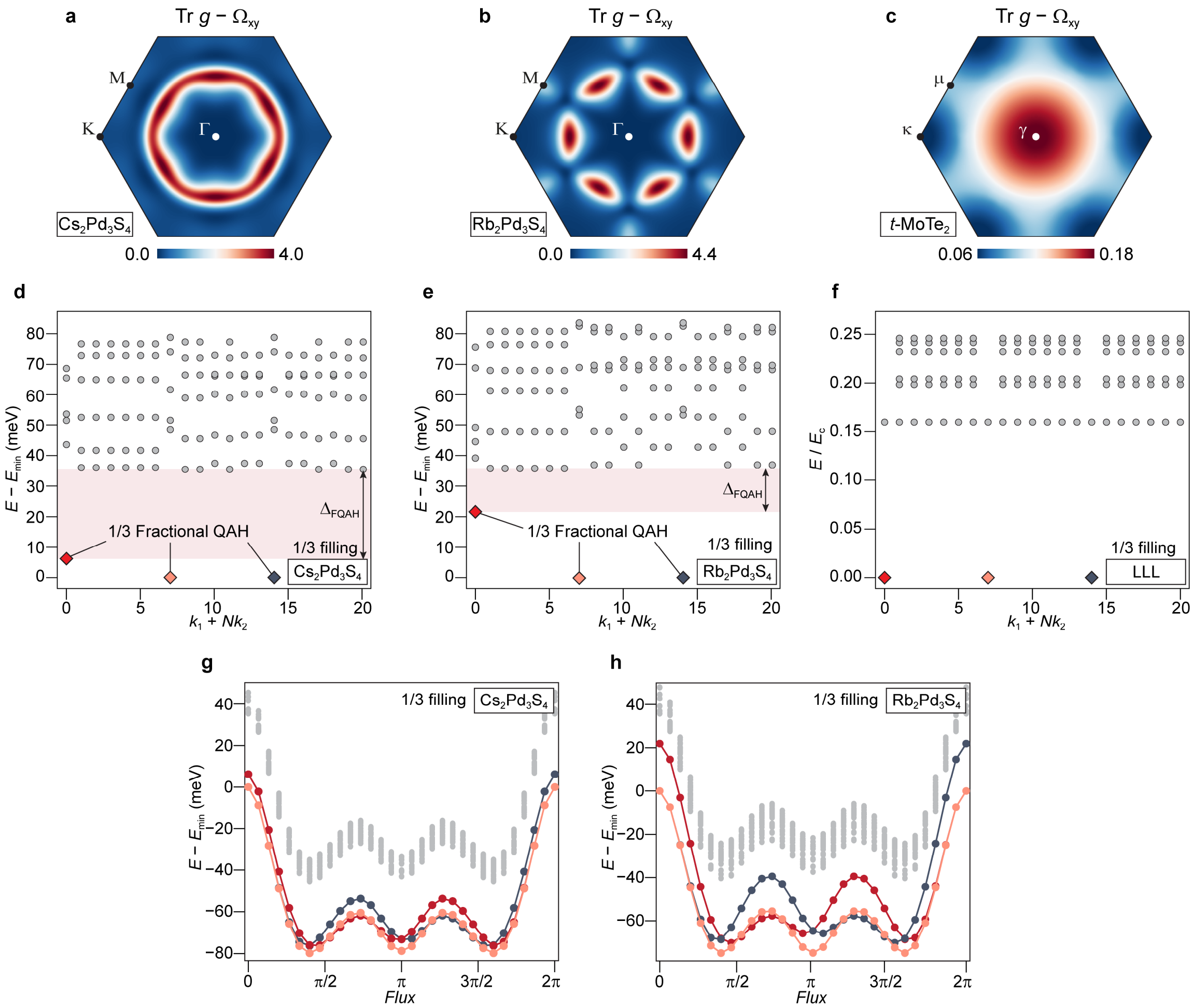

**Extended Data Figure 9 | Ideal Landau-level condition for topmost Chern bands and exact many-body spectra at fractional fillings. a–c,** The degree of similarity between a Chern band and the lowest Landau level quantified from $\frac{A_{\mathrm{BZ}}}{2\pi}\left(\mathrm{Tr}\, g_{\mathrm{k}} - \Omega_{\mathrm{k}}\right)$, where $A_{\mathrm{BZ}}$ is the Brillouin zone area, $g_{\mathrm{k}}$ is quantum metric, and $\Omega_{\mathrm{k}}$ is Berry curvature. We evaluate this quantity for the topmost Chern bands of $Cs_2Pd_3S_4$ (a), $Rb_2Pd_3S_4$ (b), and for the first Moiré band of twisted bilayer $MoTe_2$ with a twist angle of 3.89° (c). The result in c is obtained from the renormalized band structure after the Hartree-Fock calculation. While the kagome materials show maximum values that are worse, the averages over the Brillouin zone, $\frac{1}{2\pi}\int d^2k \left(\mathrm{Tr}\, g_{\mathrm{k}} - \Omega_{\mathrm{k}}\right)$, remain in the same order of magnitude– for example, 0.88 for $Cs_2Pd_3S_4$, 0.76 for $Rb_2Pd_3S_4$, and 0.11 for twisted bilayer $MoTe_2$ – setting the stage for fractional ground state in $A_2Pd_3S_4$. **d–f,** The exact many-body spectra of the 1/3-filled topmost Chern band in $Cs_2Pd_3S_4$ (d) and $Rb_2Pd_3S_4$ (e), with band mixing incorporated, at system size $N = 21$. The spectra exhibit the characteristic threefold ground-state degeneracy of the Laughlin state, mimicking that of the 1/3-filled lowest Landau level shown in f. Both the degeneracies and the momentum positions of the ground states are in agreement. We note that in d,e,g,h, energies are plotted in meV, while in f, they are expressed in units of the characteristic Coulomb interaction energy of Landau levels $E_{\mathrm{C}} = e^2/4\pi\epsilon_0 l_{\mathrm{B}}$. **g,h,** Spectral flow of the 1/3-filled states in $Cs_2Pd_3S_4$ (g) and $Rb_2Pd_3S_4$ (h) under the insertion of one flux quantum. The quasi-degenerate ground states undergo interchange and return to the original configuration only after the $6\pi$ flux insertion.

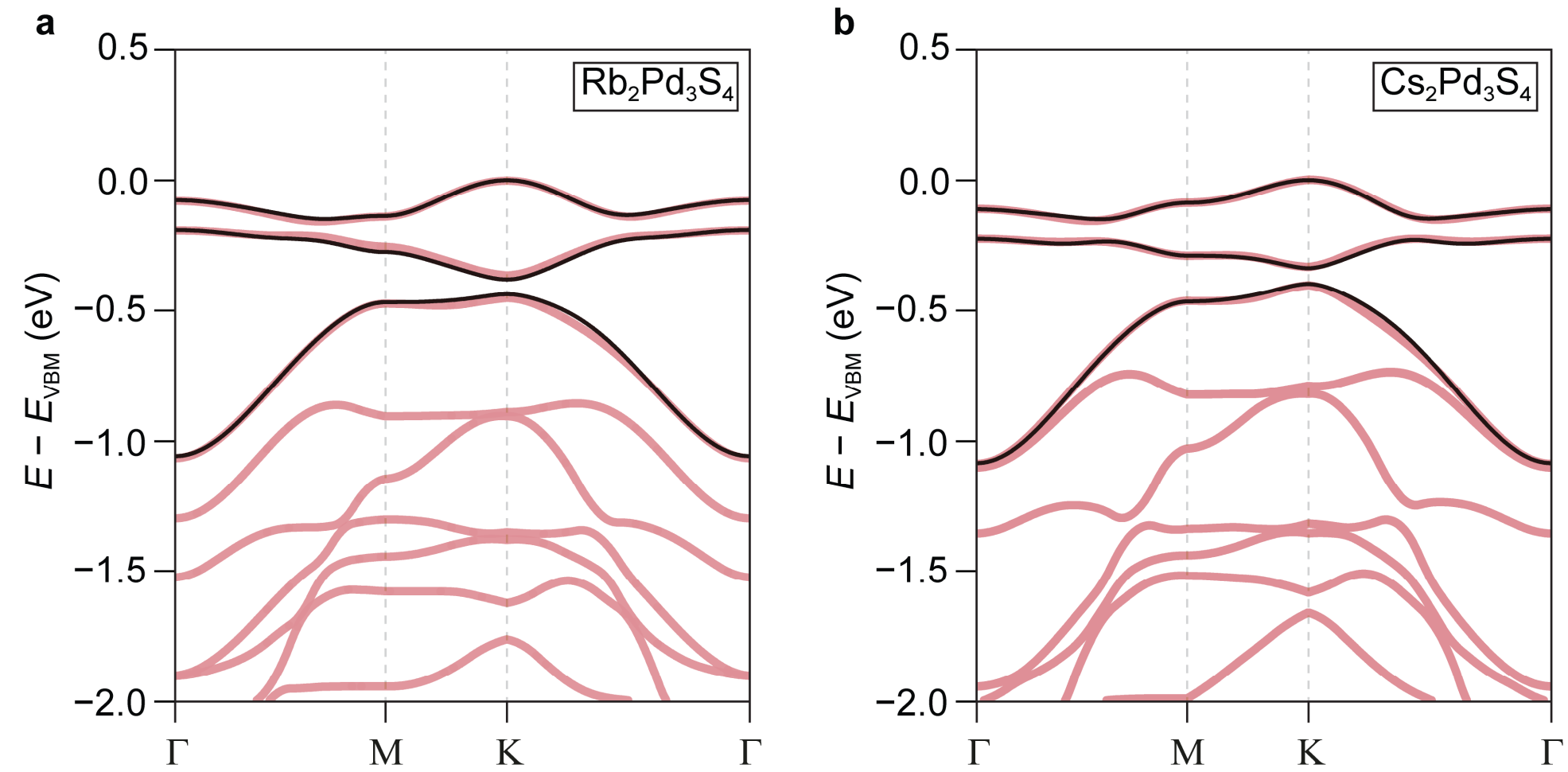

**Extended Data Figure 10 | Tight-binding band structures for $Rb_2Pd_3S_4$ and $Cs_2Pd_3S_4$. a,b,** Tight-binding dispersions (black) fit to corresponding DFT calculations (red) for $Rb_2Pd_3S_4$ and $Cs_2Pd_3S_4$, respectively. Our model is based on a single-orbital tight-binding model on the kagome lattice, including hopping terms up to the seventh-nearest neighbors and Kane-Mele spin-orbit coupling. The tight-binding model accurately reproduces the three topmost $d_{xy}$ kagome band manifolds, validating its use for modeling the key kagome physics in $A_2Pd_3S_4$.